\relax
\relax
\relax
\relax
\relax
\relax
\relax
\relax
\relax

\def\ourenc{46914}
\def\ourdec{56241}
\def\ourkeygen{156317}

\relax

\relax

\relax

\relax

\relax

\relax
\makeatletter
\documentclass[letterpaper,twocolumn,10pt]{article}
\makeatletter
\@namedef{ver@breakurl.sty}{}\@namedef{ver@cite.sty}{}\makeatother
\usepackage[hyphens]{url} \usepackage{usenix}
\usepackage[sort,numbers]{natbib}
\usepackage{hyperref}
\usepackage{threeparttable}

\date{}
\relax
\relax

\newcommand{\authormark}[1]{\textsuperscript{#1}}
\author{
{\rm
Daniel~J.~Bernstein\authormark{1,2},
Billy~Bob~Brumley\authormark{3},
Ming-Shing~Chen\authormark{2},
and
Nicola~Tuveri\authormark{3}}\\
{\tt
authorcontact-opensslntru@box.cr.yp.to}
\\
\authormark{1}Department of Computer Science, University of Illinois at Chicago, Chicago, IL 60607-7045, USA\\
\authormark{2}Ruhr University Bochum, Bochum, Germany\\
\authormark{3}Tampere University, Tampere, Finland}

\relax
\relax

\usepackage[T1]{fontenc}
\usepackage{graphicx}
\usepackage{xspace}
\usepackage{amsmath,amssymb}
\usepackage[inline]{enumitem}
\usepackage{lipsum}
\usepackage{xparse}
\usepackage{siunitx}

\setlist{noitemsep}

\usepackage{algpseudocode}
\usepackage{algorithm}
\usepackage{algorithmicx}
\usepackage{multirow}

\usepackage[available,functional,reproduced]{usenixbadges}

\newcommand{\xmm}{\ensuremath{\mathtt{xmm}}}
\newcommand{\ymm}{\ensuremath{\mathtt{ymm}}}

\newcommand{\code}[1]{\texttt{#1}}
\newcommand{\Ver}[1]{\code{#1}}

\NewDocumentCommand{\openssl}{+o} {\IfValueTF{#1}{OpenSSL~\Ver{#1}}{OpenSSL}}\NewDocumentCommand{\libcrypto}{+o} {\IfValueTF{#1}{\code{libcrypto}~\Ver{#1}}{\code{libcrypto}}}\NewDocumentCommand{\libssl}{+o} {\IfValueTF{#1}{\code{libssl}~\Ver{#1}}{\code{libssl}}}
\newcommand{\ENGINE}{\code{ENGINE}}
\newcommand{\engntru}{\code{engNTRU}}
\newcommand{\opensslLTSver}{1.1.1}
\newcommand{\opensslLTS}{\openssl[\opensslLTSver]}
\newcommand{\tlsonethree}{TLS~1.3}
\NewDocumentCommand{\libsntrup}{m} {\code{libsntrup{#1}}}\NewDocumentCommand{\sntrup}{+o} {\IfValueTF{#1}{\code{sntrup{#1}}}{\code{sntrup}}}\NewDocumentCommand{\ntruhrss}{+o} {\IfValueTF{#1}{\code{ntruhrss{#1}}}{\code{ntruhrss}}}
\newcommand{\footurl}[1]{\footnote{\url{#1}}}
\newcommand{\GLN}{\code{glib-networking}}
\newcommand{\stunnel}{\code{stunnel}}

\newcommand{\function}[1]{\code{#1()}}

\newcommand{\Encap}{\textit{\function{Encapsulate}}}
\newcommand{\Decap}{\textit{\function{Decapsulate}}}
\newcommand{\Keygen}{\textit{\function{Keygen}}}
\newcommand{\Derive}{\textit{\function{Derive}}}

\newcommand{\EVP}{\code{EVP}}
\newcommand{\evpencrypt}{\function{EVP\_PKEY\_encrypt}}
\newcommand{\evpdecrypt}{\function{EVP\_PKEY\_decrypt}}
\newcommand{\evpderive}{\function{EVP\_PKEY\_derive}}

\newcommand{\NULL}{\code{NULL}}
\newcommand{\NID}{\code{NID}}
\newcommand{\OID}{\code{OID}}

\newcommand{\msg}[1]{\texttt{#1}}
\newcommand{\ClientHello}{\msg{ClientHello}}
\newcommand{\ServerHello}{\msg{ServerHello}}
\newcommand{\keyshare}{\msg{key\_share}}
\newcommand{\Certificate}{\msg{Certificate}}
\newcommand{\CertificateVerify}{\msg{CertificateVerify}}

\newcommand{\Paragraph}[1]{\paragraph{#1.}}

\def\setof#1{\mathord{\left\lbrace{#1}\right\rbrace}}
\def\floor#1{\mathord{\left\lfloor{#1}\right\rfloor}}

\def\Z{\mathbb{Z}}

\def\R{\mathcal{R}}
\def\Rq{\R/q}

\def\thing#1{\textsf{\upshape #1}}

\newcommand{\Short}{\thing{Short}}

\newcommand{\KeyGen}{\texttt{KeyGen}}

\newcommand{\BatchKeyGen}{\texttt{BatchKeyGen}}

\newcommand{\ntrupsmall}{\textbf{small}}
\newcommand{\ntrupweight}{\textbf{weight}}

\relax
\relax

\providecommand\Paragraph{}
\relax
\relax
\relax
\renewcommand{\Paragraph}[1]{\smallbreak\noindent\textbf{#1.}\xspace}
\relax
\relax
\relax
\relax

\title{{OpenSSLNTRU}: Faster post-quantum {TLS} key exchange}

\relax

\begin{document}

\maketitle

\relax

\relax

\relax

\relax
\begin{abstract}
Google's
CECPQ1
experiment
in
2016
integrated
a
post-quantum
key-exchange
algorithm,
{\tt
newhope1024},
into
TLS
1.2.
The
Google-Cloudflare
CECPQ2
experiment
in
2019
integrated
a
more
efficient
key-exchange
algorithm,
{\tt
ntruhrss701},
into
TLS
1.3.

This
paper
revisits
the
choices
made
in
CECPQ2,
and
shows
how
to
achieve
higher
performance
for
post-quantum
key
exchange
in
TLS
1.3
using
a
higher-security
algorithm,
\sntrup[761].
Previous
work
had
indicated
that
\ntruhrss[701]
key
generation
was
much
faster
than
\sntrup[761]
key
generation,
but
this
paper
makes
\sntrup[761]
key
generation
much
faster
by
generating
a
{\it
batch\/}
of
keys
at
once.

Batch
key
generation
is
invisible
at
the
TLS
protocol
layer,
but
raises
software-engineering
questions
regarding
the
difficulty
of
integrating
batch
key
exchange
into
existing
TLS
libraries
and
applications.
This
paper
shows
that
careful
choices
of
software
layers
make
it
easy
to
integrate
fast
post-quantum
software,
including
batch
key
exchange,
into
TLS
with
minor
changes
to
TLS
libraries
and
no
changes
to
applications.

As
a
demonstration
of
feasibility,
this
paper
reports
successful
integration
of
its
fast
\sntrup[761] library,
via
a
lightly
patched
\openssl{},
into
an
unmodified
web
browser
and
an
unmodified
TLS
terminator.
This
paper
also
reports
\tlsonethree{}
handshake
benchmarks,
achieving
more
\tlsonethree{}
handshakes
per
second
than
any
software
included
in
\openssl{}.

\end{abstract}
\relax

\section{Introduction} \label{sec:intro}

The
urgency
of
upgrading
TLS
to
post-quantum
encryption
has
prompted
a
tremendous
amount
of
work.
There
were
already
69
proposals
for
post-quantum
cryptography
(PQC)
submitted
to
NIST's
Post-Quantum
Cryptography
Standardization
Project
in
2017,
including
49
proposals
for
post-quantum
encryption.
Each
proposal
included
complete
software
implementations
of
the
algorithms
for
key
generation,
encryption,
and
decryption.
Given
the
cryptographic
agility
of
TLS,
one
might
imagine
that
TLS
software
can
simply
pick
a
post-quantum
algorithm
and
use
it.
Constraints
that
make
this
more
difficult
than
it
sounds
include
the
following:
\begin{itemize}
\item
Performance:
Post-quantum
algorithms
can
send
much
more
data
than
elliptic-curve
cryptography
(ECC),
and
can
take
many
more
CPU
cycles.
Performance
plays
a
``large
role''~\cite{2020/nist-tweaks}
in
the
NIST
standardization
project.
\item
Integration:
Many
assumptions
about
how
cryptography
works
are
built
into
the
TLS
protocol
and
existing
TLS
software.
These
range
from
superficial
assumptions
about
the
sizes
of
objects
to
more
fundamental
structural
assumptions
such
as
the
reliance
of
TLS
1.3
upon
``Diffie--Hellman''---a
key-exchange
data
flow
not
provided
by
any
of
the
proposals
for
NIST
standardization.
\item
Security:
30
of
the
69
proposals
were
broken
by
the
end
of
2019~\cite{2020/bernstein-horror}.
New
attacks
continue
to
appear:
e.g.,~\cite{2020/bellare}
uses
under
a
single
second
of
CPU
time
to
break
any
ciphertext
sent
by
the
``Round2''
lattice-based
proposal.
\end{itemize}

In
July
2020,
the
NIST
project
began
its
third
round~\cite{2020/nist-8309},
selecting
4
``finalist''
and
5
``alternate''
encryption
proposals
to
consider
for
standardization
at
the
end
of
the
round
and
after
a
subsequent
round.
Meanwhile,
there
have
been
various
experiments
successfully
integrating
post-quantum
encryption
systems
into
TLS\null.
The
proposals
that
have
attracted
the
most
attention,
and
that
are
also
the
focus
of
this
paper,
are
``small''
lattice
proposals.
These
include
\begin{itemize}
\item
three
of
the
finalist
proposals
(Kyber~\cite{2019/kyber},
NTRU~\cite{2020/ntru},
and
SABER~\cite{2019/saber}),
although
NIST
says
it
will
standardize
at
most
one
of
these
three;
\item
one
of
the
alternate
proposals
(NTRU
Prime);
\item
the
{\tt
newhope1024}
algorithm~\cite{2015/alkim}
used
inside
Google's
CECPQ1
experiment
in
2016;
and
\item
the
\ntruhrss[701]
algorithm
(a
variant
of
one
of
the
algorithms
in
the
NTRU
proposal)
used
inside
the
Google-Cloudflare
CECPQ2
experiment
in
2019.
\end{itemize}
These
are
called
``small''
because
they
use
just
a
few
kilobytes
for
each
key
exchange---much
more
traffic
than
ECC,
but
much
less
than
many
other
post-quantum
proposals.

\subsection{Contributions of this paper}\label{sec:contributions}
This
paper
introduces
OpenSSLNTRU,
an
improved
integration
of
post-quantum
key
exchange
into
TLS
1.3.
OpenSSLNTRU
improves
upon
the
post-quantum
portion
of
CECPQ2
in
two
ways:
{\bf
key-exchange
performance}
and
{\bf
TLS
software
engineering}.
These
are
linked,
as
explained
below.
OpenSSLNTRU
offers
multiple
choices
of
key
sizes;
for
concreteness
we
emphasize
one
option,
\sntrup[761]~\cite{2020/ntruprime},
to
compare
to
CECPQ2's
\ntruhrss[701].

Each
of
\ntruhrss[701]/\sntrup[761]
is
a
``key-encapsulation
mechanism''
(KEM)
consisting
of
three
algorithms:
a
key-generation
algorithm
generates
a
public
key
and
a
corresponding
secret
key;
an
``encapsulation''
algorithm,
given
a
public
key,
generates
a
ciphertext
and
a
corresponding
session
key;
a
``decapsulation''
algorithm,
given
a
secret
key
and
a
ciphertext,
generates
the
corresponding
session
key.
The
key
exchange
at
the
beginning
of
a
TLS
session
involves
one
keygen,
one
enc,
and
one
dec.
Before
our
work,
both
KEMs
already
had
software
optimized
for
Intel
Haswell
using
AVX2
vector
instructions;
keygen
was
$3.03\times$
slower
for
\sntrup[761]
than
for
\ntruhrss[701],
making
total
keygen+enc+dec
$2.57\times$
slower.

\begin{table}[t]
\caption{Cryptographic features of the post-quantum components of CECPQ2
(previous
work)
and
OpenSSLNTRU
(this
paper).
Core-SVP
in
the
table
is
pre-quantum
Core-SVP
(see
\cite[Section
6]{2020/ntruprime});
post-quantum
Core-SVP
has
10\%
smaller
exponents.
See~\cite{2016/biasse}
regarding
cyclotomic
concerns.
The
\ntruhrss[701]
cycle
counts
are
from
{\tt
supercop-20210423}~\cite{2021/bernstein-ebacs}
on
{\tt
hiphop}
(Intel
Xeon
E3-1220
V3).
The
\sntrup[761]
cycle
counts
are
old$\rightarrow${\bf
new},
where
``old''
shows
the
best
\sntrup[761]
results
before
our
work
and
``{\bf
new}''
shows
results
from
this
paper's
freely
available
software;
\autoref{furthersoftware} presents the slight enc and dec speedups,
and~\autoref{sec:methods}
presents
the
large
keygen
speedup.
}\label{comparison}
\medskip
\resizebox{1.0\columnwidth}{!}{\begin{tabular}{|l|r|r|} \hline
&CECPQ2&OpenSSLNTRU\\
\noalign{\hrule}
cryptosystem&\ntruhrss[701]&\sntrup[761]\\
\noalign{\hrule}
key$+$ciphertext
bytes
&
2276
&
2197
\\
keygen
cycles
&
269191
&
814608$\rightarrow${\bf\ourkeygen}
\\
enc
cycles
&
26510
&
48892$\rightarrow${\bf\ourenc}
\\
dec
cycles
&
63375
&
59404$\rightarrow${\bf\ourdec}
\\
\noalign{\hrule}
Core-SVP
security
&
\vphantom{$2^{0^0}$}
$2^{136}$
&
$2^{153}$
\\
cyclotomic
concerns
&
yes
&
no
\\
\hline
\end{tabular}
}
\end{table}

One
can
remove
keygen
cost
by
reusing
a
key
for
many
TLS
sessions
(see
\autoref{sec:related_integrations}).
This
paper
instead
directly
addresses
the
speed
problem
with
\sntrup[761]
key
generation,
by
making
\sntrup[761]
key
generation
much
faster.
Our
\sntrup[761]
software
outperforms
the
latest
\ntruhrss[701]
software,
and
at
the
same
time
\sntrup[761]
has
a
higher
security
level
than
\ntruhrss[701].
See~\autoref{comparison}.

The
main
bottleneck
in
\sntrup[761]
key
generation
is
computation
of
certain
types
of
inverses.
This
paper
speeds
up
those
inversions
using
``Montgomery's
trick'',
the
simple
idea
of
computing
two
independent
inverses
$1/a$
and
$1/b$
as
$br$
and
$ar$
respectively,
where
$r=1/ab$.
Repeating
this
trick
converts,
e.g.,
$32$
inversions
into
$1$
inversion
plus
$93$
multiplications.

This
paper
generates
a
batch
of
$32$
independent
keys,
combining
independent
reciprocals
across
the
batch.
This
batch
size
is
large
enough
for
inversion
time
to
mostly
disappear,
and
yet
small
enough
to
avoid
creating
problems
with
latency,
cache
misses,
etc.
We
designed
new
algorithms
and
software
to
optimize
\sntrup[761]
multiplications,
since
the
multiplications
used
previously
were
``big$\times$small''
multiplications
while
Montgomery's
trick
needs
``big$\times$big''
multiplications;
see~\autoref{sec:methods}.

A
new
key
sent
through
TLS
could
have
been
generated
a
millisecond
earlier,
a
second
earlier,
or
a
minute
earlier;
this
does
not
matter
for
the
TLS
{\it
protocol}.
However,
for
TLS
{\it
software},
batching
keys
is
a
more
interesting
challenge,
for
two
reasons.
First,
key
generation
is
no
longer
a
pure
stateless
subroutine
inside
one
TLS
session,
but
rather
a
mechanism
sharing
state
across
TLS
sessions.
Second,
the
TLS
software
ecosystem
is
complicated
(and
somewhat
ossified),
with
many
different
applications
using
many
different
libraries,
so
the
same
state
change
needs
to
be
repeated
in
many
different
pieces
of
TLS
software.

To
address
the
underlying
problem,
this
paper
introduces
a
new
choice
of
software
layers
designed
to
decouple
the
fast-moving
post-quantum
software
ecosystem
from
the
TLS
software
ecosystem.
The
point
of
these
layers
is
that
optimization
of
post-quantum
software
does
not
have
to
worry
about
any
of
the
complications
of
TLS
software,
and
vice
versa.
As
a
case
study
demonstrating
the
applicability
of
these
layers,
this
paper
describes
successful
integration
of
its
new
\sntrup[761]
library,
including
batch
key
generation,
into
an
existing
web
browser
communicating
with
an
existing
TLS
terminator,
using
OpenSSL
on
both
ends.
This
demo
involves
no
changes
to
the
web
browser,
no
changes
to
the
TLS
terminator,
and
very
few
changes
to
OpenSSL.

The
integration
of
OpenSSLNTRU
into
TLS
means
that,
beyond
microbenchmarks,
we
can
and
do
measure
full
TLS
handshake
performance.
The
bottom
line
is
that,
in
a
controlled
and
reproducible
end-to-end
lab
experiment,
\sntrup[761] completes more sessions per second than
commonly
deployed
pre-quantum
NIST
P-256,
and
even
completes
more
sessions
per
second
than
commonly
deployed
pre-quantum
X25519
(see
\autoref{sec:e2e_experiment:benchmark}).
This
remains
true
even
when
we
replace
\sntrup[761] with higher-security \sntrup[857].

\section{Background}
\label{sec:backgrounds}

\subsection{Polynomial rings in NTRU Prime}\label{sec:notation}

Streamlined
NTRU
Prime~\cite{2020/ntruprime},
abbreviated~\sntrup[],
uses
arithmetic
in
finite
rings
$\R/3=(\Z/3)[x]/(x^p-x-1)$
and
$\R/q=(\Z/q)[x]/(x^p-x-1)$,
where
$\R=\Z[x]/(x^p-x-1)$.
The
parameters
$p,q$
are
chosen
so
that
$\R/q$
is
a
field.

$\Short$
means
the
set
of
polynomials
in
$\R$
that
are
\ntrupsmall,
meaning
all
coefficients
in
$\setof{-1,0,1}$,
and
\ntrupweight~$w$,
meaning
that
exactly
$w$
coefficients
are
nonzero,
where
$w$
is
another
parameter.
The
parameters
$(p,q,w)$
are
$(653,4621,288)$,
$(761,4591,286)$,
$(857,5167,322)$
for
the
KEMs
\sntrup[653],
\sntrup[761],
\sntrup[857]
respectively.

\subsection{Montgomery's trick for batch inversion}
\label{sec:mont}

In
this
section,
we
review
Montgomery's
trick
for
batch
inversion~\cite{Montgomery1987SpeedingTP}
as
applied
to
many
inputs.
The
algorithm
\texttt{batchInv}
takes
$n$
elements
$(
a_1,
a_2,
\ldots
,
a_n
)$
in
a
ring,
and
outputs
their
multiplicative
inverses
$
(a_1^{-1},
a_2^{-1}
,
\ldots,
a_n^{-1})
$.
Montgomery's
trick
for
batch
inversion
proceeds
as
follows:
\begin{enumerate}
\item Let $b_1 = a_1$ and compute $b_i = a_i \cdot b_{i-1}$ for $i$ in $(2,\ldots,n)$.
After
$n-1$
multiplications,
we
obtain
\relax
\[ (b_1,b_2,\ldots,b_n) = ( a_1 , a_1\cdot a_2 , a_1 \cdot a_2 \cdot a_3 , \ldots , \Pi_{i=1}^{n} a_i ) \enspace . \]
\relax

\item Compute the single multiplicative inverse
\relax
\[ t_n = b_n^{-1} = ( \Pi_{i=1}^{n} a_i )^{-1} \enspace. \]
\relax
\item Compute $ c_i = t_i \cdot b_{i-1} $ and $ t_{i-1} = t_{i} \cdot a_{i}$ for $i$ in $(n,\ldots,2)$.
After
$2
n
-
2
$
multiplications,
we
have
two
lists
\relax
\[
\begin{aligned}
(c_n,\ldots,c_2)
 &
=
(a_n^{-1},\ldots,a_2^{-1})
 \enspace
\hbox{and}
\\
(t_{n-1},
\ldots,
t_2
,t_1)
&
=(
(\Pi_{i=1}^{n-1}
a_i)^{-1}
,
\ldots
,
(a_1\cdot
a_2)^{-1}
,
a_1^{-1}
)
\enspace
.
\end{aligned}
\]
\relax
\item Output $(a_1^{-1}, a_2^{-1} , \ldots, a_n^{-1})$.
\end{enumerate}
In
summary,
the
algorithm
uses
$3n
-
3$
multiplications
and
one
inversion
to
compute
$n$
inverses.

\subsection{NTT-based multiplication}
\label{sec::FFT:multiplication}

This
section
reviews
techniques
for
polynomial
multiplication
commonly
used
in
lattice-based
cryptography.
We
adopt
terminology
from~\cite{2001/bernstein}.

The
number
theoretic
transform
(NTT)
algorithm
maps
an
element
in
a
polynomial
ring
into
values
by
lifting
the
ring
element
to
a
polynomial
and
evaluating
the
polynomial
on
a
particular
set.
An
NTT-based
multiplication
algorithm
applies
NTTs
to
two
input
elements
in
the
polynomial
ring,
performs
component-wise
multiplication
for
the
transformed
values,
and
applies
an
inverse
NTT,
converting
the
multiplied
values
back
to
the
product
in
the
same
form
of
inputs.

Computing
a
size-$n$
NTT,
where
$n$
is
a
power
of
$2$,
comprises
$\log_2
n$
stages
of
the
\textit{radix-2
FFT
trick}.
Given
a
polynomial
ring
$(\Z/q)[x]/(x^{n}-b^2)$
where
$b
\in
\Z/q$,
the
FFT
trick
maps
elements
in
$
(\Z/q)[x]/(x^{n}-b^2)
$
to
$
(
(\Z/q)[x]/(x^{n/2}-b)
)
\times
(
(\Z/q)[x]/(x^{n/2}
+
b)
)
$.
Due
to
the
Chinese
remainder
theorem
(CRT),
the
mapping
is
invertible
when
$2b$
is
invertible.
Specifically,
let
$f
=
f_0
+
f_1
x
+
\cdots
+
f_{n-1}
x
^{n-1}
 \in
(\Z/q)[x]/(x^{n}-b^2)
$.
The
trick
maps
$f$
to
\begin{align*}
&
(f
\bmod
(x^{n/2}+b)
\enspace
,
f
\bmod
(x^{n/2}-b)
)
 \\
=
&
(
(f_0
-
b
f_{n/2})
+
\cdots
+
(f_{n/2-1}
-
b
f_{n-1})
x^{n/2-1}
 ,
\\
&
 (f_0
+
b
f_{n/2})
+
\cdots
+
(f_{n/2-1}
+
b
f_{n-1})
x^{n/2-1}
 )
\end{align*}
with
$n$
multiplications
by
$b$,
$n/2$
additions,
and
$n/2$
subtractions.
Setting
$b
=
1$,
by
recursively
applying
the
FFT
trick,
an
NTT
transforms
$f$
into
a
list
$\hat{f}
=
(\hat{f}_0,\ldots,
\hat{f}_{j},
\ldots,\hat{f}_{n-1})
\in
(\Z/q)^n$
where
$\hat{f}_j
=
f
\bmod
(x-\psi^{j})
=
\sum_{i=0}^{n-1}
f_i
\psi^{ij}$,
and
$\psi
\in
\Z/q$
is
a
primitive
$n$-th
root
of
unity,
i.e.,
$\psi^{n/2}
=
-1$.

When
$\Z/q$
lacks
appropriate
roots
of
unity,
Sch\"{o}nhage's
trick~\cite{77/Schoennhage}
manufactures
them
by
introducing
an
intermediate
polynomial
ring.
Given
$
f
\in
(\Z/q)[x]/(x^{2mn}
-
1)$,
the
trick
first
introduces
a
new
variable
$y
=
x^m$
and
maps
$f$
from
 $(\Z/q)[x]/(x^{2mn}
-
1)
$
to
$
((\Z/q)[x][y]/(y^{2n}-1))/(x^m-y)
$.
Then,
it
lifts
$f$
to
$(\Z/q)[x][y]/(y^{2n}-1)$,
which
is
a
polynomial
in
variable
$y$
with
coefficients
in
$(\Z/q)[x]$.
Since
the
coefficients
of
$f$
are
polynomials
with
degree
less
than
$m$,
it
is
safe
to
map
them
to
$
(\Z/q)[x]/(x^{2m}+1)$
such
that
coefficient
multiplication
needs
no
reduction
by
$x^{2m}+1$.
Now
$x
\in
(\Z/q)[x]/(x^{2m}+1)
$
is
a
primitive
$4m$-th
root
of
unity,
since
$
x^{2m}
=
-1$.

Nussbaumer's
trick~\cite{80/Nussbaumer}
is
another
method
to
manufacture
roots
of
unity.
Given
$
f
\in
(\Z/q)[x]/(x^{2mn}
-
1)$,
the
trick
maps
$f$
to
$((\Z/q)[y]/(y^{2n}+1))[x]/(x^m-y)$,
lifts
to
$((\Z/q)[y]/(y^{2n}+1))[x]$,
and
maps
to
$((\Z/q)[y]/(y^{2n}+1))[x]/(x^{2n}-1)$
for
$n\ge
m$.
As
noted
in~\cite{2001/bernstein},
Nussbaumer's
trick
sometimes
uses
slightly
smaller
ring
extensions
than
Sch\"onhage's
trick,
but
Sch\"onhage's
trick
is
more
cache-friendly,
since
it
uses
contiguous
data
in
$x$.

\subsection{The AVX2 instruction set}

Since
NIST
specified
Intel
Haswell
CPU
as
its
highest
priority
platform
for
performance
evaluation~\cite{NIST:guideline},
we
optimize
\sntrup[]
for
the
Haswell
architecture
in
this
work.

Specifically,
we
target
the
Advanced
Vector
Extensions
2
(AVX2)
instruction
set.
AVX
is
a
single-instruction-multiple-data
(SIMD)
instruction
set
in
modern
(decade
or
less)
x86
CPUs.
It
provides
sixteen
256-bit
\ymm{}
registers;
each
\ymm{}
register
splits
into
two
128-bit
\xmm{}
lanes.
The
instruction
set
treats
data
in
\ymm{}
registers
as
lanes
(independent
partitions)
of
32$\times$8-bit,
16$\times$16-bit,
8$\times$32-bit,
etc.;
every
instruction
operates
simultaneously
on
the
partitioned
data
in
the
\ymm{}
registers.
In
2013,
the
Haswell
architecture
extended
AVX
to
AVX2
for
enhanced
integer
operations.

\subsection{Related works}\label{sec:related}

\subsubsection{NTT-based multiplication in other PQC finalists}

Among
the
lattice
based
KEM
of
NIST's
finalists,
Kyber~\cite{2019/kyber}
operates
in
a
radix-2
NTT
friendly
polynomial
ring
and
implements
NTT-based
multiplication
in
the
proposal.
SABER~\cite{2019/saber}
and
NTRU~\cite{2020/ntru}
operate
in
polynomial
rings
with
a
power-of-two
modulus
which
are
considered
NTT-unfriendly.
The
earlier
implementations
of
two
schemes
used
a
combination
of
Toom-4
and
Karatsuba
based
polynomial
multiplication.

Recently,~\cite{2021/Chung}
showed
that
NTT-based
multiplication
outperforms
previous
Toom-Cook
multiplication
for
implementing
NTT-unfriendly
SABER
and
most
parameters
of
NTRU.
To
use
NTT-based
multiplication
in
an
NTT-unfriendly
ring,
they
raise
the
coefficients
to
a
combination
of
several
NTT-friendly
polynomial
rings,
perform
several
NTT-based
multiplications,
and
map
back
to
original
ring
with
CRT.
For
NTRU
on
the
AVX2
platform,
they
reported
significant
improvement
for
parameters
with
polynomials
of
degree
greater
then
700.
For
Saber,
they
also
reported
a
pronounced
performance
gain
although
the
degree
of
polynomials
are
only
255.
It
is
because
the
matrix-vector
multiplication
allows
them
to
save
the
input
NTT
transforms
for
the
elements
in
the
common
vector
which
performs
inner
products
with
different
rows
in
the
matrix.

\subsubsection{Integrating cryptographic primitives}\label{sec:related_integrations}

Related
to
OpenSSLNTRU,
several
previous
works
studied
integrations
between
post-quantum
implementation
and
real
world
applications
and
protocols.

The
Open~Quantum~Safe~(OQS)
project~\cite{DBLP:conf/sacrypt/StebilaM16}
includes
a
library
of
quantum-resistant
cryptographic
algorithms,
and
prototype
integrations
into
protocols
and
applications.
It
also
includes
(and
requires)
a
fork
of
the
\openssl{}
project.
Conversely,
in
our
contribution
we
apply
a
minimal
patchset,
striving
to
maintain
API
and
ABI
compatibility
with
the
\openssl{} version available to the end-users. This avoids the need of
recompiling
existing
applications
to
benefit
from
the
new
library
capabilities.
While
\cite{DBLP:conf/sacrypt/StebilaM16}
focused
primarily
on
key
agreement,
the
OQS
OpenSSL
fork
does
also
support
signatures
and
certificates
using
post-quantum
algorithms,
and
their
negotiation
in
TLS.
See
\cite{DBLP:conf/pqcrypto/PaquinST20}
for
a
study,
conducted
using
OQS,
benchmarking
post-quantum
TLS
authentication.
We
also
note
that
the
end-to-end
experiment
we
present
in
this
paper
is
limited
to
one
candidate
and
two
sets
of
parameters
(\sntrup[761]
and
\sntrup[857]),
while
the
OQS
project
provides
implementations
for
all
finalists.

Similarly,
the
PQClean
project~\cite{github/pqclean}
collects
a
number
of
implementations
for
the
candidates.
However,
it
does
not
aim
to
include
integration
into
higher-level
applications
or
protocols.

CECPQ2
actually
included
two
experiments:
CECPQ2a
used
\ntruhrss[701],
while
CECPQ2b
used
an
isogeny-based
proposal.
Compared
to
\ntruhrss[701],
the
isogeny-based
proposal
had
smaller
keys
and
smaller
ciphertexts,
but
used
much
more
CPU
time,
so
it
outperformed
CECPQ2a
only
on
the
slowest
network
connections.

In
general,
the
importance
of
a
few
kilobytes
depends
on
the
network
speed
and
on
how
often
the
application
creates
new
TLS
sessions.
A
typical
multi-megabyte
web
page
is
unlikely
to
notice
a
few
kilobytes,
even
if
it
retrieves
resources
from
several
TLS
servers.
A
session
that
encrypts
a
single
DNS
query
is
handling
far
less
data,
making
the
performance
of
session
establishment
much
more
important.
Similar
comments
apply
to
CPU
time.

\begin{figure}[t]
\centering
\includegraphics[width=1.0\columnwidth]{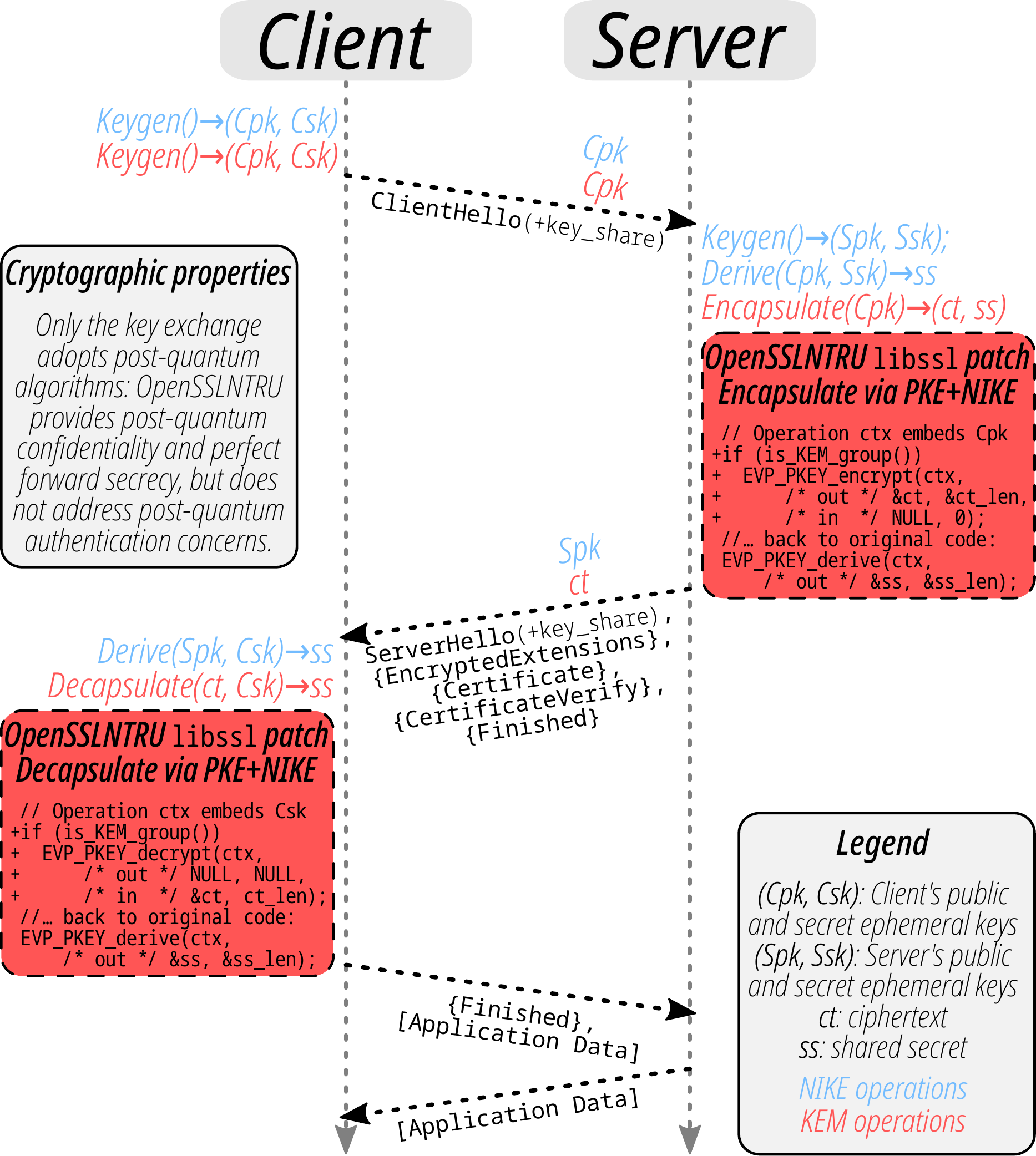}
\caption{Overview of a full \tlsonethree{} handshake.
In
OpenSSLNTRU,
the
traditional
NIKE
operations
are
replaced
with
KEM
operations.
Besides
dedicated
\code{NamedGroup}
codepoints,
this
is
transparent
to
\tlsonethree{}
messages
as
\keyshare{}
payloads
are
opaque.
As
\opensslLTS{}
does
not
offer
an
API
for
KEM
operations,
our
patch
uses
the
described
PKE+NIKE
workaround
when
one
of
the
supported
KEM
groups
is
negotiated.
}\label{fig:tls13diagram}
\end{figure}

\citet*{DBLP:conf/ccs/SchwabeSW20} present an alternative to the
\tlsonethree{} handshake to solve both key exchange and authentication
using
post-quantum
KEM.
In
contrast,
for
our
experiment
we
aimed
at
full
compatibility
with
the
\tlsonethree{}
ecosystem,
focusing
exclusively
on
the
key
exchange.
This
ensures
post-quantum
confidentiality,
but
does
not
address
the
post-quantum
authentication
concerns.
Therefore,
showcasing
how
at
the
protocol
level
our
experiment
does
not
alter
the
\tlsonethree{}
message
flow,
in
\autoref{fig:tls13diagram}
we
only
highlight
the
cryptographic
operations
and
material
involved
in
the
key
exchange---carried
in
the
\ClientHello{}
and
\ServerHello{}
messages---while
keys
and
signatures
used
for
authentication---as
part
of
the
\Certificate{} and \CertificateVerify{} messages---do not address
post-quantum
concerns.

Our
approach
to
\openssl{}
integration
via
an
\ENGINE{}
module
is
based
on
the
methodology
suggested
in
\cite{DBLP:conf/secdev/TuveriB19}, where the authors instantiated
\code{libsuola}.
In
this
context,
an
\ENGINE{}
module
is
a
a
dynamically
loadable
module.
Using
a
dedicated
API,
such
a
module
is
capable
of
injecting
new
algorithms
or
overriding
existing
ones.
The
implementations
it
provides
can
be
backed
by
a
hardware
device,
or
be
entirely
software
based.
Our
new
\ENGINE{},
\engntru{},
builds
upon
\code{libbecc}~\cite{DBLP:conf/latincrypt/BrumleyHSTV19},
which
is
itself
derived
from
\code{libsuola}.
Both
previous
works
applied
the
\ENGINE{}
framework
to
integrate
alternative
ECC
implementations.
The
latter
is
particularly
close
to
\engntru{},
as
it
also
featured
a
transparent
mechanism
to
handle
batch
key
generation.
\autoref{sec:engntru} details how \engntru{} evolved from these works
and
the
unique
features
it
introduces.

\citet*{SB02:batching} integrated RSA batching to improve SSL
handshake
performance
already
in
2001.
However,
their
methodology
required
integrating
changes
directly
in
the
server
application.
In
contrast,
OpenSSLNTRU
acts
on
the
middleware
level,
transparent
to
client
and
server
applications.

\Paragraph{Comparison table}Based
on
the
previous
discussions
in
this
section,
\autoref{tab:tlscompare}
compares
select
TLS
integration
experiments
regarding
post-quantum
algorithms.

The
``Hybrid''
criterion
tracks
approaches
that
simultaneously
protect
the
key
agreement
with
``traditional''
(usually
ECC)
and
post-quantum
encryption
(see,
e.g.,
\cite{ietf-tls-hybrid-design,DBLP:conf/pqcrypto/BindelBFGS19}).
This
paper
does
not
make
recommendations
for
or
against
hybrids;
our
performance
and
software-engineering
contributions
are
equally
applicable
to
hybrid
and
non-hybrid
scenarios.
\autoref{fig:tls13diagram} illustrates how
any
NIKE
system
can
be
transformed
into
an
equivalent
KEM
construction;
a
protocol
that
supports
key
exchange
via
KEM
can
support
hybrid
handshakes
by
simply
composing
two
or
more
underlying
KEMs
to
obtain
a
hybrid
KEM.

The
``PFS''
criterion
tracks
approaches
that
provide
the
traditional
notion
of
Perfect
Forward
Secrecy
w.r.t.\ the
key
agreement
phase
of
the
handshake.
Different
experiments
take
post-quantum
security
into
consideration
at
different
cryptosystem
components,
tracked
by
the
``key
agreement''
and
``authentication''
criteria.
The
latter
comes
with
the
caveat
that
the
extent
to
which
PQ
authentication
is
achieved
is
inherently
limited
by
access
to
a
fully
post-quantum
Public
Key
Infrastructure
(PKI).
In
the
specific
case
of
the
Internet
Web
PKI,
client
and
server
need
to
share
a
chain
of
certificates
up
to
a
common
root
of
trust,
entirely
signed
with
PQ
algorithms.
Some
experiments
require
breaking
changes
to
the
\tlsonethree{}
message
flow,
depicted
in
\autoref{fig:tls13diagram};
``compatibility''
tracks
this
criterion.
Lastly,
our
work
is
the
only
experiment
we
are
aware
of
that
achieves
ABI
compatibility
(``Binary'')
to
easily
integrate
into
the
existing
software
ecosystem.

\begin{table}[t]
\caption{Comparison of select TLS integration experiments.
}
\medskip
\label{tab:tlscompare}
\resizebox{1.0\columnwidth}{!}{\begin{threeparttable}
\begin{tabular}{|l|r|r|r|r|r|}
\hline
&
OQS

&
CECPQ2
&
KEMTLS
&
OpenSSLNTRU
\\
\hline
Hybrid\tnote{1}
\vphantom{$2^{0^0}$}
&
opt.\tnote{7}

&
yes
 
 &
no

&
no
 \\
PFS\tnote{2}

&
yes

&
yes
 
 &
yes
 
 &
yes
\\
PQ-sec.\ key
agmt.\tnote{3}

 &
yes

&
yes
 
 &
yes
 
 &
yes
\\
PQ-sec.\ auth.\tnote{4}

 &
opt.

 &
no

&
yes
 
 &
no
 \\
\tlsonethree{} compat.\tnote{5}      & yes               & yes    & no     & yes \\
Binary
compat.\tnote{6}

 &
no

 &
no

&
no

&
yes
\\
\hline
\end{tabular}
\begin{tablenotes}

\item [1]
Key-agreement
uses
ECC
\emph{and}
post-quantum
encryption.
\item [2]
Key-agreement
provides
Perfect
Forward
Secrecy.
\item [3]
Post-quantum
security
over
TLS
key
agreement.
\item [4]
Post-quantum
security
over
TLS
authentication,
inherently
limited
by
access
to
PQ
PKI.
\item [5]
Requires
no
breaking
changes
to
the
\tlsonethree{}
message
flow.
\item [6]
ABI
compatible,
to
easily
integrate
into
the
existing
software
ecosystem.
\item [7]
\cite{DBLP:conf/sacrypt/StebilaM16} presents experimental results
for
both
post-quantum
and
hybrid
KEMs.
Using
the
OQS
fork
of
\openssl{},
the
choice
of
supported
KEMs
and
order
of
preference
is
left
to
developers
and
system
administrators.
\end{tablenotes}
\end{threeparttable}
}
\end{table}

\section{Batch key generation for \code{sntrup}}
\label{sec:methods}

This
section
presents
batch
key
generation
for
\sntrup[]
and
its
optimization.
\autoref{sec:batchkeygen} shows the batch key generation algorithm with Montgomery's inversion-batching trick.
\autoref{sec:polymul:z3} and \autoref{sec:polymul:zq} present our polynomial multiplication and its optimization in $(\Z/3)[x]$
and
$(\Z/q)[x]$,
respectively.
\autoref{sec:benchmark:batchkeygen} shows the benchmark results.

\subsection{Batch key generation}
\label{sec:batchkeygen}

The
\sntrup[]
key
generation
algorithm
$\KeyGen$
outputs
an
\sntrup[]
key
pair.
It
proceeds
as
follows:
\begin{enumerate}
\item
Generate
a
uniform
random
small
element
$g\in\R$.
\emph{Repeat this step until $g$ is invertible in $\R/3$.}
\item
Compute
$1/g$
in
$\R/3$.
\item
Generate
a
uniform
random
$f\in\Short$.
\item
Compute
$h=
g
/(3f)
$
in
$\Rq$.
\item
Output
$(h,(f,1/g))$
where
$h$
is
the
public
key
and
$(f,1/g)
\in
\Short
\times
\R/3$
is
the
secret
key.
\end{enumerate}

\autoref{alg:batchkeygen}~(\BatchKeyGen) batches \sntrup[] key generation.
We
use
two
lists
for
storing
$n$
batches
of
$g\in\R$
and
$f\in\Short$,
then
process
the
$n$
batches
of
computation
in
one
subroutine.
The
key
idea
is
to
replace
the
$2n$
inversions
by
two
$\texttt{batchInv}$
for
$\R/3$
and
$\Rq$,
respectively.
As
seen
in
\autoref{sec:mont},
\texttt{batchInv}
turns
$n$
inversions
into
$3n-3$
multiplications
and
one
inversion.
Considering
performance,
ring
multiplication
then
becomes
the
critical
part.
Hence,
\autoref{sec:polymul:z3}
and
\autoref{sec:polymul:zq}
present
optimized
ring
multiplication
implementations,
used
in
$\texttt{batchInv}$.

Another
difference
is
the
invertibility
check
in
$\R/3$
for
the
element
$g$.
Previous
NTRU
Prime
software
checks
invertibility
as
a
side
effect
of
computing
$1/g$
with
a
constant-time
algorithm~\cite{BY19:extgcd}
for
extended
GCD\null.
Calling
$\texttt{batchInv}$
removes
this
side
effect
and
requires
a
preliminary
check
for
invertibility
of
each
$g$.
In~\autoref{sec:iszero}
we
optimize
an
\texttt{isInvertible}
subroutine
for
this
test.

\begin{algorithm}
\caption{\BatchKeyGen}
\label{alg:batchkeygen}
\hspace*{\algorithmicindent} \textbf{Input :} an integer $n$ \\
\hspace*{\algorithmicindent} \textbf{Output:} $n$ key pairs of \sntrup[]
\begin{algorithmic}[1]

\State $ G \gets \left[ \cdot \right] $  \Comment{an empty list}
\State $ F \gets \left[ \cdot \right] $
\While{ $ \code{len}(G) < n $}
\State $ g \xleftarrow{\$} \R/3 $    \Comment{$\$$: uniform random}
\State \textbf{if} not \code{isInvertible}( $g$ ) \textbf{: continue }
\State $ f \xleftarrow{\$} \Short $
\State $ G\code{.append}(g) $
\State $ F\code{.append}(f) $
\EndWhile
\State $ \bar{G} \gets \code{batchInv}(G) $
\State $ \bar{F} \gets \code{batchInv}( \left[ 3\cdot f \textbf{ for } f \in F \right] ) $
\State $ H \gets \left[ g \cdot \bar{f} \in \R/q \textbf{ for } g \in G , \bar{f} \in \bar{F}  \right] $
\State \textbf{return} $ \left[ ( h , (f,\bar{g}) ) \textbf{ for } h \in H, f \in F, \bar{g} \in \bar{G} \right] $
\end{algorithmic}
\end{algorithm}

\subsubsection{Invertibility check for elements in $\R/3$}
\label{sec:iszero}

At
a
high
level,
we
check
the
invertibility
of
an
element
$
g
\in
\R/3$
by
computing
its
remainder
of
division
by
the
irreducible
factors
of
$
x^p
-
x
-
1$
modulo
$3$,
as
suggested
in~\cite[p.~8]{2020/ntruprime}.
This
section
optimizes
this
computation.

For
convenience,
we
always
lift
the
ring
element
$
g
$
to
its
polynomial
form
$
g
\in
(\Z/3)[x]$
in
this
section.
In
a
nutshell,
if
$g
\bmod
f_i
=
0$
for
any
factor
$f_i$
of
$
x^p
-
x
-
1$,
then
$g$
is
not
invertible
in
$\R/3$.

We
calculate
the
remainder
of
$g
\bmod
f_i
$
with
Barrett
reduction~\cite{HAC96}.
Suppose
the
polynomial
$x^p-x
-
1
\in
(\Z/3)[x]$
has
$m$
irreducible
factors
$(f_1,\ldots,f_m)$,
i.e.,
$x^p
-
x
-
1
=
\Pi_{i=1}^{m}
f_i$.
Given
a
polynomial
$g
\in
(\Z/3)[x]$
and
$p
>
\deg(g)
>
\deg(f_i)
$,
we
calculate
the
reminder
$r
=
g
\bmod
f_i$
as
follows.
In
the
\textit{pre-computation}
step,
choose
$D_g
>
\deg(g)
$
and
$
D_{f_i}
>
\deg(f_i)$,
and
calculate
$q_x$
as
the
quotient
of
the
division
$
x^{D_g}
/
f_i
$,
i.e.,
$
q_x
=
\floor{
x^{D_g}
/
f_i
}
$,
where
the
floor
function
$\floor{
\cdot
}$
removes
the
negative-degree
terms
from
a
series.
In
the
\textit{online}
step,
compute
$
h
=
\floor{
g
\cdot
q_x
/
x^{D_g}
}
 =
\floor{
g
/
f_i
}
$,
i.e.,
the
quotient
of
the
division
$
g
\cdot
q_x
/
x^{D_g
}
$.
Finally,
return
the
remainder
$r
=
g
-
h
\cdot
f_i
$.
We
show
this
gives
the
correct
$r$
in
\autoref{sec:proofmod3}.

Some
observations
about
the
degree
of
polynomials
help
to
accelerate
the
computation.
While
computing
$
h
=
\floor{
g
\cdot
q_x
/
x^{D_g}
}
$,
we
compute
only
terms
with
degree
in
the
interval
$[0,D_f)$,
since
$r
=
g-
h
\cdot
f_i$
uses
terms
exclusively
from
this
interval
for
$\deg(r)
<
\deg(f_i)$.

In
the
case
of
$\sntrup[761]$,
the
polynomial
$f
=
x^{761}
-
x
-
1
\in
(\Z/3)[x]$
has
three
factors,
with
degrees
$\deg(f_1)
=
19$,
$\deg(f_2)
=
60$,
and
$\deg(f_3)
=
682$,
respectively.
We
choose
$D_{f_1}
=
32$,
$D_{f_2}
=
64$,
and
$D_{g}
=
768$
for
computing
$g
\bmod
f_0$
and
$g
\bmod
f_1$.
For
computing
$g
\bmod
f_3$,
we
note
the
pre-computed
quotient
$
q_x
=
\floor{
x^{768}
/
(x^{682}+\cdots)
}
$
satisfies
$\deg(q_x)=88$.
Hence,
the
multiplication
$
h
=
\floor{
g
\cdot
q_x
/
x^{768}
}
$
involves
$\deg(g)=768$
and
$\deg(q_x)=88$
polynomials.
By
partitioning
the
longer
polynomial
into
several
shorter
segments,
we
perform
the
multiplication
by
several
polynomial
multiplications
of
length
equal
to
the
shorter
polynomial
(less
than
$128$).
Therefore,
to
check
invertibility,
we
use
polynomial
multiplications
in
$(\Z/3)[x]$
with
lengths
in
$\{32,64,128\}$.

\subsection{Polynomial multiplication in $(\Z/3)[x]$}
\label{sec:polymul:z3}

In
this
section,
we
describe
our
multiplication
in
$(\Z/3)[x]$
for
$\sntrup[]$,
and
its
optimization
in
the
AVX2
instruction
set.

Based
on
the
polynomial
lengths,
we
implement
polynomial
multiplication
with
different
algorithms.
We
build
a
$16
\times
16$
polynomial
multiplier
as
a
building
block
for
schoolbook
multiplication.
We
then
use
Karatsuba
to
build
longer
multipliers,
such
as
$32
\times
32$,
$64
\times
64$,
and
further
$
2^i
\times
2^i$.
For
$3\cdot
256
\times
3\cdot
256
$
multiplications,
we
start
from
Bernstein's
5-way
recursive
algorithm~\cite{djb09:karatsuba}
for
$(\Z/2)[x]$
and
optimize
the
same
idea
for
$(\Z/3)[x]$.

\subsubsection{Base polynomial multiplier}
\label{sec::16x16}

For
representing
$(\Z/3)[x]$
polynomials,
we
adjust
the
values
of
coefficients
to
unsigned
form
and
store
polynomials
as
byte
arrays,
with
one
coefficient
per
byte.
For
example,
we
store
the
polynomial
$a_0
+
\cdots
+
a_{15}x^{15}
\in
(\Z/3)[x]$
as
a
byte
array
$(a_0,a_1,\ldots,a_{15})$
in
a
16-byte
\xmm{}
register.

Besides
a
byte
array,
we
can
view
a
polynomial
as
an
integer
by
translating
the
monomial
$x
=
256$.
For
example,
a
degree-3
polynomial
$a_0
+
a_1
x
+
a_2
x^2
+
a_3
x^3$
maps
to
the
32-bit
integer
$
a_0
+
a_1\cdot
2^8
+
a_2
\cdot
2^{16}
+
a_3
\cdot
2^{24}
 $.

In
this
32-bit
format,
we
can
perform
a
$4
\times
4
\rightarrow
8$
polynomial
multiplication
using
a
$32
\times
32
\rightarrow
64$
\textit{integer}
multiplication,
taking
care
to
control
the
coefficient
values.
While
calculating
the
polynomial
product
$
(a_0
+
a_1
x
+
a_2
x^2
+
a_3
x^3)
\cdot
(b_0
+
b_1
x
+
b_2
x^2
+
b_3
x^3)$
with
a
$32
\times
32
\rightarrow
64$
integer
multiplication,
if
all
coefficients
$a_i,
b_i
\in
\left\{
0,1,2
\right\}
$,
a
term's
maximum
possible
value
is
$
\sum_{i+j=3}
a_i
b_j
x^3
\leq
16$,
fitting
in
a
byte.
Hence,
we
use
$4
\times
4$
polynomial
multiplication
(i.e.,
$32
\times
32
\rightarrow
64$
integer
multiplication),
as
our
building
block
to
implement
$16
\times
16$
polynomial
multiplication
with
the
schoolbook
algorithm.

\subsubsection{Multiplying polynomials of length $3n$}
\label{sec:refine:karatsuba}

This
section
reduces
a
multiplication
of
$3n$-coefficient
polynomials
in
$(\Z/3)[x]$
to
$5$
multiplications
of
${\approx}n$-coefficient
polynomials,
while
optimizing
the
number
of
additions
using
techniques
analogous
to
Bernstein's
optimizations~\cite{djb09:karatsuba}
for
$(\Z/2)[x]$.
This
section
also
streamlines
the
computation
for
${\le}(3n-1)$-coefficient
polynomials,
as
in
\sntrup[].

Take
two
polynomials
$F_0
+
F_1
t
+
F_2
t^2
$
and
$G_0
+
G_1
t
+
G_2
t^2$
in
$(\Z/3)[x]$,
where
$\deg(F_i)
<
n$,
$\deg(G_i)
<
n$,
and
$t
=
x^n$.
Their
product
$
H
=
H_0
+
H_1
t
+
H_2
t^2
+
H_3
t^3
+
H_4
t^4
$
can
be
reconstructed
by
the
projective
Lagrange
interpolation
formula
\relax
\[
\begin{aligned}
H
 =
&
H(0)
\frac{(t-1)(t+1)(t-x)}{x}
 +
H(1)
\frac{t(t+1)(t-x)}{x-1}
\\
&
+
H(-1)
\frac{t(t-1)(t-x)}{x+1}
+
 H(x)
\frac{t(t-1)(t+1)}{x(x-1)(x+1)}
\\
&
+
H(\infty)
t(t-1)(t+1)(t-x)
 \enspace
.
\end{aligned}
\]
Here
\[
\begin{aligned}
H(0)
 &=
F_0
\cdot
G_0,
\\
H(1)
 &=
(F_0
+
F_1
+
F_2)\cdot
(G_0
+
G_1
+
G_2),
\\
H(-1)
&=
(F_0
-
F_1
+
F_2)\cdot
(G_0
-
G_1
+
G_2),
\\
H(x)
 &=
(F_0
+
F_1
x
+
F_2
x^2)\cdot
(G_0
+
G_1
x
+
G_2
x^2),\hbox{
and}
\\
H(\infty)
&=
F_2
\cdot
G_2
\end{aligned}
\]
\relax
are
the
only
five
polynomial
multiplications
in
the
algorithm.
These
polynomials
expand
from
$n$
to
$2n$
terms,
except
$H(x)$.

\relax

$H$
simplifies
to
\begin{equation}
\label{eq:h}
\begin{aligned}
H
&
=
 H(0)
-
\left[
U
+
(H(1)-H(-1))
\right]
\cdot
t
 \\
&
-
\left[
H(0)
+
(H(1)+H(-1))
+
H(\infty)
\right]
\cdot
t^2
\\
&
+
U
\cdot
t^3
 +
H(\infty)
\cdot
t^4
\enspace
,
\end{aligned}
\end{equation}
where
$
U
=
V
+
{H(0)}/{x}
-
H(\infty)
\cdot
x
$
and
\[
V
=
\frac{
(
H(1)+H(-1))\cdot
x
+
(H(1)-H(-1))
+
H(x)/x
 }{
x^2-1
}
\enspace.
\]
\relax

There
are
two
tricky
issues
while
computing
$V$.
First,
$\deg(H(x))
\leq
2n
+
2$,
introducing
extra
complexity
since
all
other
polynomials
have
degree
less
than
$2n$.
By
requiring
$\deg(F_2)
\le
n-2$
and
$\deg(G_2)
\le
n-2$,
we
force
$\deg(H(x))
\leq
2n$.
Since
$H(x)$
is
only
used
as
$H(x)/x$
in
$V$,
we
can
always
process
polynomials
with
degree
less
than
$2n$.

The
other
issue
concerns
computing
divisions
by
$x^2-1$
in
$(\Z/3)[x]$.
Since
long
division
is
a
sequential
process
and
not
efficient
in
SIMD
settings,
we
now
present
a
divide-and-conquer
method
for
it.

\subsubsection{Division by $x^2-1$ on $(\Z/3)[x]$}

Dividing
a
polynomial
$f$
by
$x^2-1$
means
producing
a
representation
of
$f
=
q
\cdot
(x^2-1)
+
r
$,
where
$q$
and
$r
=
r_1
x
+
r_0$
are
the
quotient
and
remainder,
respectively.
Assume
that
we
have
recursively
divided
two
$2m$-coefficient
polynomials
$f$
and
$g$
by
$x^2-1$,
obtaining
$f
=
q
\cdot
(x^2-1)
+
r$
and
$g
=
s
\cdot
(x^2-1)
+
t$.
Then
\relax
\begin{equation*}
\begin{aligned}
r
\cdot
x^{2m}
&=
(
r
x^{2m-2}
+
r
 x^{2m-4}
+
r
x^{2m-6}+
\cdots
+
r)
(x^2-1)
+
r
 \enspace
,
\end{aligned}
\end{equation*}
\relax
so
the
result
of
dividing
$f
\cdot
x^{2m}
+
g$
by
$(x^2-1)$
is
\relax
\begin{equation}
\label{eq:divxsq1}
\begin{aligned}
f
\cdot
x^{2m}
+
g
&
=
\left[
q
\cdot
x^{2m}
+
r
\cdot
x^{2m-2}
\right]
(x^2-1)
 \\
&
+
(
s
+
r
x^{2m-4}
+
\cdots
+
r
)
(x^2-1)
+
(t
+
r)
\enspace
.
\end{aligned}
\end{equation}
\relax
We
carry
out
these
divisions
in
place
as
follows:
recursively
overwrite
the
array
of
$f$
coefficients
with
$q$
and
$r$,
recursively
overwrite
the
array
of
$g$
coefficients
with
$s$
and
$t$,
and
then
simply
add
the
lowest
two
coefficients
from
the
$f$
array
into
every
coefficient
pair
in
the
$g$
array.

Because
the
recursive
computations
for
$f$
and
$g$
are
independent,
this
computation
parallelizes.
The
overall
parallel
computation
for
dividing
a
length-$n$
polynomial
by
$x^2-1$,
assuming
$n=2^l$,
proceeds
as
follows.
The
computation
comprises
$l-1$
steps.
The
first
step
splits
the
polynomial
into
$n/4$
separate
sub-polynomials;
each
sub-polynomial
has
degree
less
than
four.
We
divide
a
length-four
sub-polynomial
by
$x^2-1$
by
adding
two
coefficients
of
higher
degrees
to
the
lower
two
coefficients.
We
perform
these
divisions
in
parallel.
In
each
subsequent
step,
we
double
the
sub-polynomial
sizes,
and
divide
sub-polynomials
by
$x^2-1$
by
adding
two
coefficients
of
lower
degree
from
the
higher
degree
parts
to
the
lower
parts
of
the
polynomials
as
in
\autoref{eq:divxsq1}.
Since
each
step
performs
$n/2$
additions,
the
whole
computation
costs
$n(\log_2(n)-1)/2$
additions.

\subsubsection{AVX2 optimization for the $\R/3$ multiplier}

Since
we
use
integer
arithmetic
for
$\Z/3$
and
integers
grow,
we
must
control
the
values
to
prevent
overflow.
From
the
AVX2
instruction
set,
we
use
the
$\texttt{vpshufb}$
instruction
to
reduce
the
values.
The
instruction
reads
the
lower
nibbles
as
indexes
from
single-byte
lanes
of
a
register,
then
replaces
the
lane
values
with
those
from
a
16-entry
table,
using
the
four-bit
indexes.
Thus,
we
use
$\texttt{vpshufb}$
to
reduce
integers
in
$[0,16)$
to
integers
in
$[0,3)$.
We
also
reduce
adjacent
nibbles
by
moving
them
to
lower
positions
using
bit-shift
instructions.

Our
software
for
$16\times
16$
polynomial
multiplication
actually
performs
two
independent
$16
\times
16$
multiplications
in
the
two
\xmm{}
lanes
of
\ymm{}
registers,
respectively.
The
approach
avoids
the
high
latency
for
moving
data
between
different
\xmm{}
lanes
in
Haswell
CPUs
(see~\cite[p.~237]{fog2021instruction}
for
the
\code{vperm2i128},
\code{vextracti128},
and
\code{vinserti128}
instructions).
Specifically,
our
AVX2
multiplier
takes
two
\ymm{}
registers
as
input
and
outputs
products
in
two
\ymm{}
registers.
A
\ymm{}
register
comprises
two
polynomials
$(a,c)$
where
$a,
c
\in
(\Z/3)[x]$
are
stored
in
different
\xmm{}
lanes.
Given
two
\ymm{}
inputs
$(a,c)$
and
$(b,d)$,
the
multiplier
outputs
$(ab_{l},cd_{l})$
and
$(ab_h,cd_h)$
in
two
\ymm{}
registers,
where
$
a\cdot
b
=
ab_l
+
ab_h
\cdot
x^{16}
$
and
$
c
\cdot
d
=
cd_l
+
cd_h
\cdot
x^{16}
$.
Thus,
we
avoid
the
data
exchange
between
\xmm{}
lanes.

\subsection{Polynomial multiplication in $(\Z/q)[x]$}
\label{sec:polymul:zq}

\Paragraph{Problem description and related multiplication}
While
applying
NTT-based
multiplication,
NTRU
Prime
faces
two
issues.
First,
recalling
\autoref{sec:notation},
NTRU
Prime
works
on
the
polynomial
ring
$\R/q
=
(\Z/q)[x]/(x^p-x-1)$
where
$x^p-x-1$
is
irreducible
in
$(\Z/q)[x]$;
hence,
there
is
no
way
to
apply
FFT
tricks
on
the
ring.
The
standard
workaround
is
to
lift
ring
elements
in
$\R/q$
to
$(\Z/q)[x]$,
and
multiply
the
lifted
polynomials
with
an
NTT-based
multiplication
in
$(\Z/q)[x]/(x^N-1)$
where
$N
\ge
2p
$.
Since
two
input
polynomials
have
degree
less
than
$p$,
their
product
will
not
overflow
the
degree
$N$.
After
the
polynomial
multiplication,
the
product
is
reduced
with
a
division
by
$x^p-x-1$
for
the
result
in
$\R/q$.

Secondly,
$q$
from
the
NTRU
Prime
parameter
set
is
not
a
radix-2
NTT
friendly
prime.
For
example,
$q=4591$
in
\sntrup[761],
and
since
$4591
-
1
=
2
\cdot
3^3
\cdot
5
\cdot
17
$,
no
simple
root
of
unity
is
available
for
recursive
radix-2
FFT
tricks.
\citet*{2020/alkim-polymult} presented a non-radix-2 NTT implementation
on
$(\Z/{4591})[x]/(x^{1530}-1)$
for
embedded
systems.
They
performed
radix-3,
radix-5,
and
radix-17
NTT
stages
in
their
NTT.
We
instead
use
a
radix-2
algorithm
that
efficiently
utilizes
the
full
\ymm{}
registers
in
the
Haswell
architecture.

The
fastest
Haswell
\sntrup[]
software
before
our
work
dealt
with
the
radix-2-NTT-unfriendly
$q$
by
lifting
the
coefficients
to
$\Z$
and
then
multiplying
in
$(\Z/{7681})[x]$
and
$(\Z/{10753})[x]$.
Both
$7681$
and
$10753$
are
NTT-friendly.
This
suffices
for
``big$\times$small''
multiplications
for
all
specified
NTRU
Prime
parameters:
one
input
is
a
small
element
of
$\R/q$,
coefficients
in
$\setof{
-1,0,1
}$;
the
maximum
coefficient
of
a
``big$\times$small''
product
is
below
$7681\cdot
10753/2$
in
absolute
value.

However,
Montgomery's
trick
involves
general
``big$\times$big''
multiplications
in
$\R/q$.
Even
if
each
coefficient
for,
e.g.,
$q=4591$
is
fully
reduced
to
the
range
$[-2295,2295]$,
the
product
here
can
have
coefficients
as
large
as
$2295\cdot
2295\cdot
761>7681\cdot
10753$.
One
way
to
handle
these
multiplications
would
be
to
use
more
NTT-based
multiplications
over
small
moduli,
for
example
multiplying
in
$(\Z/{7681})[x]$
and
$(\Z/{10753})[x]$
and
$(\Z/{12289})[x]$,
but
this
means
50\%
more
NTTs,
plus
extra
reductions
since
$12289$
is
larger
than
$10753$.
We
take
a
different
approach
described
below.

\Paragraph{Our polynomial multiplication}
In
this
section,
we
present
a
multiplication
for
polynomials
in
$(\Z/q)[x]$
with
degree
less
than
$1024$.
We
first
map
polynomials
to
$(\Z/q)[x]/(x^{2048}
-
1)
$.
Rather
than
switching
from
$q$
to
an
NTT-friendly
prime,
we
use
Sch\"{o}nhage's
trick
(\autoref{sec::FFT:multiplication})
to
manufacture
roots
of
unity
for
radix-2
NTTs.

Specifically,
define
$K$
as
the
ring
$(\Z/q)[x]/(x^{64}+1)$.
We
map
$(\Z/q)[x]/(x^{2048}
-
1)$
to
$((\Z/q)[y]/(y^{64}
-
1)
)[x]/(x^{32}-y)$,
lift
to
$
(\Z/q)[x][y]/(y^{64}
-
1)
$,
and
then
map
to
$
K[y]/(y^{64}-1)
$.
Each
$32$
consecutive
terms
of
a
polynomial
in
$(\Z/q)[x]$
are
thus
viewed
as
an
element
of
$K$.
We
segment
the
original
polynomial
of
$1024$
terms
in
$x$
into
$32$
elements
in
$K$,
associating
each
element
in
$K$
to
a
new
indeterminate
$y$
with
different
degrees.
The
remaining
problem
is
to
multiply
elements
of
the
ring
$
K[y]/(y^{64}-1)$.

We
use
NTTs
to
multiply
in
$K[y]/(y^{64}-1)$,
using
$x$
as
a
primitive
$128$-th
root
of
unity
in
$K$.
NTT-based
multiplication
applies
two
NTTs
for
the
input
polynomials,
performs
component-wise
multiplication
for
the
transformed
values,
and
applies
one
inverse
NTT
for
the
final
product.
Each
NTT
converts
one
input
element
in
$K[y]/(y^{64}-1)$
into
64
elements
in
$K$,
using
additions,
subtractions,
and
multiplications
by
powers
of
$x$.
Multiplication
by
a
power
of
$x$
simply
raises
the
degree
of
the
polynomial
in
$(\Z/q)[x]$,
and
then
replaces
$x^{64+i}$
by
$-x^i$,
using
negations
without
any
multiplications
in
$\Z/q$.

After
transforming
the
input
polynomials
into
a
list
of
elements
in
$K$,
we
perform
the
component-wise
multiplication
for
the
transformed
vectors.
The
problem
now
is
to
multiply
two
elements
of
$K=(\Z/q)[x]/(x^{64}+1)$.

We
use
Nussbaumer's
trick
(\autoref{sec::FFT:multiplication})
to
manufacture
further
roots
of
unity:
map
$K$
to
$((\Z/q)[y]/(y^8+1))[x]/(x^8-y)$,
lift
to
$
((\Z/q)[y]/(y^8+1))[x]
$,
and
map
to
$
((\Z/q)[y]/(y^8+1))[x]/(x^{16}-1)
$.
The
polynomial
ring
$(\Z/q)[y]/(y^8+1)$
supports
a
radix-2
NTT
of
size
16
with
a
primitive
root
of
unity
$y$.
Since
the
polynomials
are
short,
we
choose
Karatsuba's
algorithm
for
component-wise
multiplication
in
$(\Z/q)[y]/(y^8+1)$.
We
use
Montgomery
multiplication~\cite{85/Montgomery}
to
calculate
modular
products
in
$\Z/q$.

For
\sntrup[761]
and
\sntrup[653],
the
input
polynomials
have
degree
less
than
$
768$,
so
we
truncate
some
computations
in
the
NTT
algorithm:
we
apply
NTT
on
the
ring
$
K
[y]/((y^{32}+1)(y^{16}-1))$
instead
of
the
original
$K[y]/(y^{64}-1)$.
We
map
the
input
polynomials
to
degree-$24$
polynomials
in
$K[y]$,
and
calculate
the
product
with
a
truncated
inverse
NTT
of
48
values.
Our
NTT
sizes
are
within
$18$\%,
$1$\%,
and
$20$\%
of
optimal
for
$653$,
$761$,
and
$857$
respectively;
further
truncation
is
possible
at
the
expense
of
some
complication
in
the
calculations.

\Paragraph{AVX2 optimization for the $\R/q$ multiplier}
Since
the
component-wise
multiplication
step
comprises
48
or
64
multiplications
on
$K$,
we
perform
the
multiplications
simultaneously
in
different
16-bit
lanes
of
\ymm{}
registers.
Our
software
stores
the
first
$\Z/q$
coefficient
of
16
elements
in
$K$
in
a
\ymm{}
register,
stores
their
second
coefficients
in
a
second
register,
and
so
on.
In
this
way,
we
avoid
data
movement
between
the
16-bit
lanes
inside
a
\ymm{}
register.

To
apply
this
optimization,
we
first
rearrange
the
coefficients
of
a
polynomial
to
different
registers
with
a
$16
\times
16$
matrix
transposition.
Given
sixteen
degree-15
polynomials
$
(a_0^{(0)}+a_1^{(0)}
x
+
\cdots
+
a_{15}^{(0)}
x^{15})
,
\cdots
,
(a_0^{(15)}
+
\cdots
+
a_{15}^{(15)}
x^{15})
$,
data
in
$(
\ldots
)$
represents
one
\ymm{}
register,
and
we
treat
a
polynomial
in
one
\ymm{}
register
as
a
row
of
a
$16
\times
16$
matrix.
Transposing
this
matrix
rearranges
the
data
to
$
(a_0^{(0)}
,
\cdots
,
a_{0}^{(15)}
)
,
\cdots
,
(a_{15}^{(15)}
,
\cdots
,
a_{15}^{(15)}
)
$.
Thus,
we
can
fetch
a
specific
coefficient
by
accessing
its
corresponding
\ymm{}
register,
while
parallelizing
16
polynomial
multiplications
for
the
transposed
data.

We
use
the
method
in~\cite{2012/hackersdelight}
for
matrix
transposition.
The
technique
transposes
a
$2
\times
2$
matrix
by
swapping
its
two
off-diagonal
components.
For
transposing
matrices
with
larger
dimensions,
e.g.\ $4
\times
4$,
it
first
swaps
data
between
two
$2
\times
2$
off-diagonal
sub-matrices,
and
then
performs
matrix
transpose
for
all
its
four
sub-matrices.

\subsection{Microbenchmarks: arithmetic}
\label{sec:benchmark:batchkeygen}

We
benchmark
our
implementation
on
an
Intel
Xeon
E3-1275
v3
(Haswell),
running
at
3.5
GHz,
with
Turbo
Boost
disabled.
The
numbers
reported
in
this
section
are
medians
of
3
to
63
measurements,
depending
on
the
latency
of
the
operation
under
measurement.
We
omit
benchmarks
here
for
\sntrup[653]
because
it
actually
uses
the
same
multiplier
as
\sntrup[761].

\Paragraph{Benchmarks for $\R/3$}
We
compare
cycle
counts
for
$\R/3$
multiplication
between
our
implementation
and
the
best
previous
\sntrup[]
implementation,
\texttt{round2}
in
\cite{2021/bernstein-ebacs},
in
the
following
table.
\begin{center}
\begin{tabular}{ | c | c | r |  }
\hline
Parameter

 &
Implementation

&
Cycles
 
\\
\hline
\multirow{3}{*}{ \sntrup[761] }
&
this
work
(\autoref{sec:polymul:z3})

&
$
8183
$
\\
 \cline{2-3}
&
this
work
(NTT,
\autoref{furthersoftware})
&
$
8827
$
\\
 \cline{2-3}
&
NTRUP
\texttt{round2}
(NTT,
\cite{2021/bernstein-ebacs})

 &
$
9290
$
\\
 \hline
\multirow{3}{*}{ \sntrup[857] }
&
this
work
(\autoref{sec:polymul:z3})

&
$
12840$
\\
 \cline{2-3}
&
this
work
(NTT,
\autoref{furthersoftware})
&
$
12533$
\\
 \cline{2-3}
&
NTRUP
\texttt{round2}
(NTT,
\cite{2021/bernstein-ebacs})

 &
$
12887$
\\
 \hline

\end{tabular}
\end{center}
The
best
results
are
from
our
our
Karatsuba-based
polynomial
multiplication
for
smaller
parameters,
and
from
our
NTT
improvements
for
larger
parameters.

Another
question
is
the
efficiency
of
Montgomery's
trick
for
inversion
in
$\R/3$.
Recall
that,
roughly,
the
trick
replaces
one
multiplicative
inversion
by
three
ring
multiplications,
one
amortized
ring
inversion,
and
one
check
for
zero
divisors.
We
show
benchmarks
of
these
operations
in
the
following
table.
\begin{center}
\begin{tabular}{ | c | c | r |  }
\hline
Parameter

 &
Operation

 &
Cycles
 
\\
\hline
\multirow{3}{*}{ \sntrup[653] }
&
Ring
inversion

&
$
95025
 $
 
\\
 \cline{2-3}
&
Invertibility
check
 
 &
$
22553
 $
 
\\
 \cline{2-3}
&
Ring
multiplication
 
 &
$
 8063
 $
 
\\
 \hline
\multirow{3}{*}{ \sntrup[761] }
&
Ring
inversion

&
$
114011
$
 
\\
 \cline{2-3}
&
Invertibility
check
 
 &
$
 9668
 $
 
\\
 \cline{2-3}
&
Ring
multiplication
 
 &
$
 8183
 $
 
\\
 \hline
\multirow{3}{*}{ \sntrup[857] }
&
Ring
inversion

&
$
160071
$
 
\\
 \cline{2-3}
&
Invertibility
check
 
 &
$
 12496
$
 
\\
 \cline{2-3}
&
Ring
multiplication
 
 &
$
 12533
$
 
\\
 \hline

\end{tabular}
\end{center}
We
can
see
the
cost
of
three
multiplications
and
one
invertibility
check
is
less
than
half
of
a
single
inversion
in
$\R/3$.
It
is
clear
that
batch
inversion
costs
less
than
pure
ring
inversion,
even
for
the
smallest
possible
batch
size
of
two.

\Paragraph{Benchmarks for $\R/q$}
The
following
table
shows
the
cycle
counts
of
big$\times$big
multiplication
and
big$\times$small
multiplication
in
$\R/q$,
comparing
with
the
previous
best
software~\cite{2021/bernstein-ebacs}.

\medskip

\noindent
\resizebox{1.0\columnwidth}{!}{\begin{tabular}{ | c | c | r |  }
\hline
Parameter

 &
Implementation
 &
 
Cycles
 
\\
\hline
\multirow{3}{*}{\sntrup[761]}
&
this
work
(\autoref{sec:polymul:zq}),
big$\times$big

 &
$25113
 $
 
\\
 \cline{2-3}
&
this
work
(\autoref{furthersoftware}),
big$\times$small
 
 &
$16992
 $
 
\\
 \cline{2-3}
&
NTRUP
\texttt{round2}~\cite{2021/bernstein-ebacs},
big$\times$small
 
 &
$18080
 $
 
\\
 \hline
\multirow{3}{*}{\sntrup[857]}
&
this
work
(\autoref{sec:polymul:zq}),
big$\times$big

 &
$32265
 $
 
\\
 \cline{2-3}
&
this
work
(\autoref{furthersoftware}),
big$\times$small
 
 &
$24667
 $
 
\\
 \cline{2-3}
&
NTRUP
\texttt{round2}~\cite{2021/bernstein-ebacs},
big$\times$small
 
 &
$25846
 $
 
\\
 \hline
\end{tabular}
}

\medskip

\noindent
The
results
show
the
absolute
cycle
count
of
big$\times$big
is
larger
than
big$\times$small
multiplication.
To
evaluate
the
efficiency
of
big$\times$big
multiplication,
consider
if
we
extend
the
big$\times$small
multiplication
to
big$\times$big
multiplication,
by
applying
more
internal
NTT
multiplications.
It
will
result
in
multiplications
of
roughly
$3/2$
times
the
current
cycle
counts,
i.e.,
slower
than
big$\times$big
multiplication
presented
in
this
work.

Since
big$\times$small
multiplication
is
faster
than
big$\times$big,
we
use
the
former
as
much
as
possible
in
$\texttt{batchInv}$
for
$\R/q$.
Recall
that
Montgomery's
trick
for
batch
inversion
replaces
one
inversion
in
$\R/q$
by
roughly
three
ring
multiplications
and
one
amortized
ring
inversion.
From
the
$\texttt{batchInv}$
algorithm
in
\autoref{sec:mont},
we
can
see
the
three
ring
multiplications
are
$a_i
\cdot
b_{i-1}$,
$a_i
\cdot
t_i$,
and
$t_i
\cdot
b_{i-1}$.
Since
the
input
$a_i$
is
a
small
element,
it
turns
out
that
only
the
last
is
big$\times$big
multiplication.
Since
the
costs
for
inverting
one
element
in
$\R/q$
are
$576989$,
$785909$,
and
$973318$
cycles
for
\sntrup[653],
\sntrup[761],
and
\sntrup[857],
respectively,
the
cost
of
two
big$\times$small
and
one
big$\times$big
multiplication
is
clearly
much
less
than
one
inversion
operation.

\begin{figure}[t]
\centering
\includegraphics[width=1.0\columnwidth]{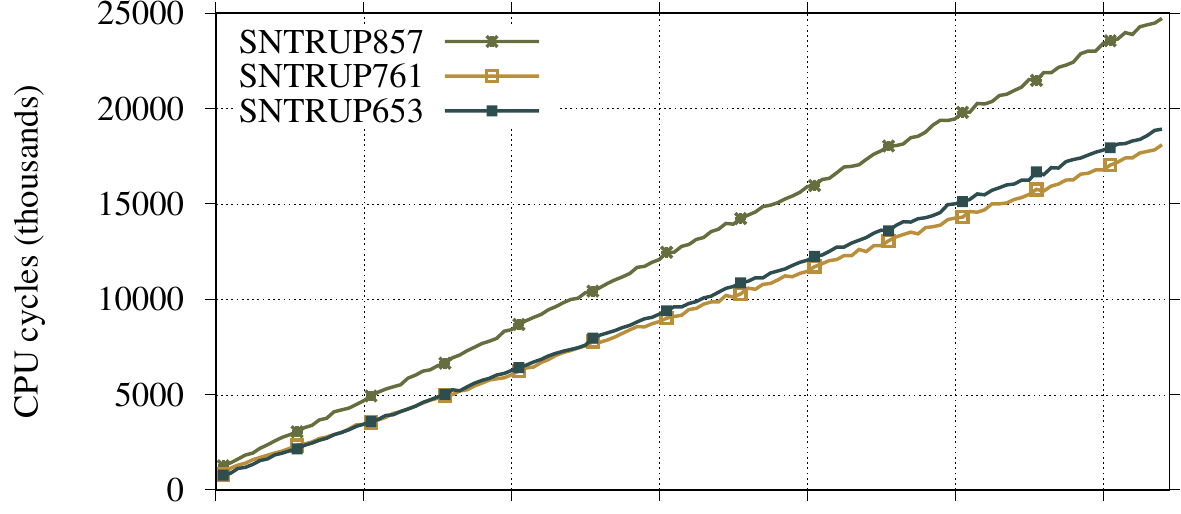}\\
\includegraphics[width=1.0\columnwidth]{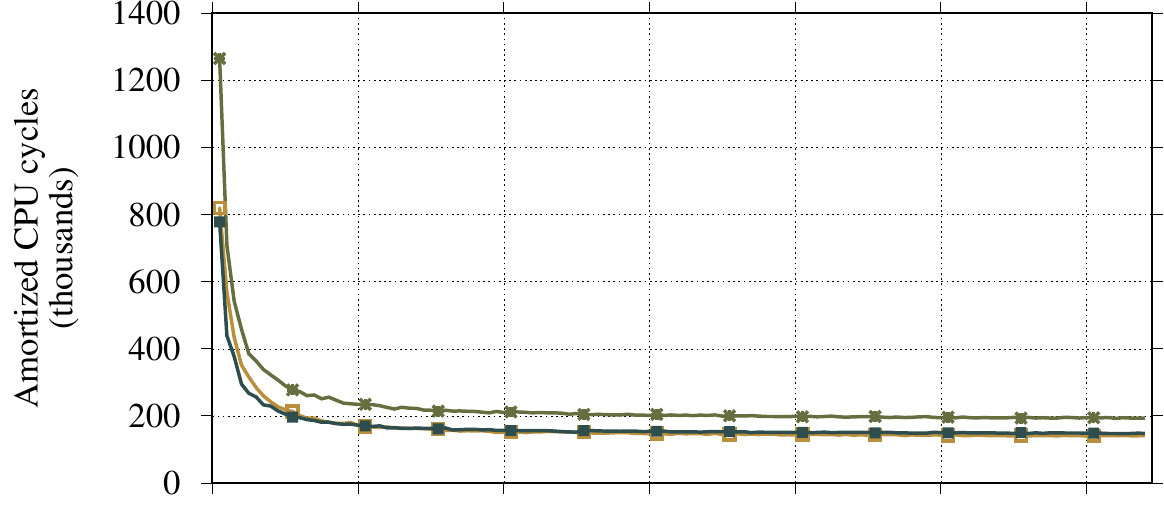}\\
\includegraphics[width=1.0\columnwidth]{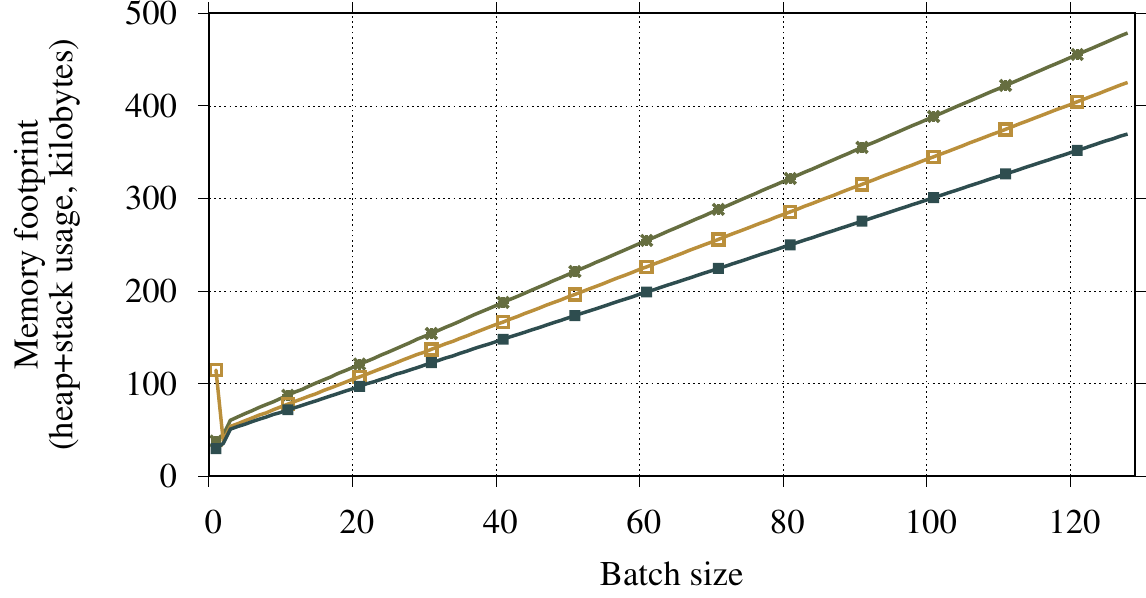}
\caption{$\BatchKeyGen$ metrics regarding various batch sizes ($n$).
Top:
full
batch
cost
in
CPU
cycles.
Middle:
amortized
cost
in
CPU
cycles,
dividing
by
$n$.
Bottom:
memory
footprint,
i.e.,
heap+stack
usage,
in
kilobytes.}
\label{fig:p_result}
\end{figure}

\Paragraph{Benchmarks for batch key generation}
We
show
the
benchmark
results
for
batch
key
generation
(\BatchKeyGen)
in
\autoref{fig:p_result}.
See
also~\autoref{tab::keygen:performance:2}.

The
figure
shows
how
increasing
$n$,
the
key
generation
batch
size,
amortizes
the
ring
inversion
cost.
Generating
a
few
dozen
keys
at
once
already
produces
most
of
the
throughput
benefit:
for
example,
generating
$n=32$
keys
takes
a
total
of
1.4
milliseconds
for
\sntrup[761]
at
3.5GHz.
Generating
$n=128$
keys
takes
a
total
of
5.2
milliseconds
for
\sntrup[761]
at
3.5GHz,
about
10\%
better
throughput
than
$n=32$.

We
adopt
$\BatchKeyGen$
with
batch
size
$n=32$
in
our
library,
resulting
in
{\bfseries
\ourkeygen{}
Haswell
cycles
per
key}.

\section{New TLS software layering}\label{sec:e2e_experiment}

At
the
application
level,
the
goals
of
our
end-to-end
experiment
are
to
demonstrate
how
the
new
results
can
be
deployed
in
real-world
conditions,
transparently
for
the
end
users,
and
meet
the
performance
constraints
of
ubiquitous
systems.
For
this
reason,
we
developed
patches
for
\opensslLTS{}
to
support
post-quantum
key
exchange
for
\tlsonethree{}
connections.
We
designed
our
patches
so
that
any
existing
application
built
on
top
of
\opensslLTS{}
can
transparently
benefit
from
the
PQC
enhancements
with
no
changes,
as
the
patched
version
of
\openssl{}
retains
API/ABI
compatibility
with
the
original
version
and
acts
as
a
drop-in
replacement.
This
works
for
any
application
dynamically
linking
against
\libssl{}
as
the
backend
to
establish
\tlsonethree{}
connections.
Among
them,
for
our
demonstration,
we
picked
a
web
browser\footnote
{GNOME
Web
(a.k.a.\ \texttt{epiphany})
---
\url{https://wiki.gnome.org/Apps/Web}},
a
custom
application
(\code{tls\_timer},
described
later),
and
a
TLS
proxy.\footnote
{\texttt{stunnel}
---
\url{https://www.stunnel.org/}}
After
installing
our
patched
version
of
\openssl{},
users
can
establish
secure
and
fast
post-quantum
TLS
connections.

\autoref{sec:ossl_primer} provides relevant technical background regarding
the
\openssl{}
software
architecture.
The
rest
of
this
section
describes,
with
more
detail,
our
work
to
achieve
the
goals
of
our
experiment,
and
provides
rationale
for
the
most
relevant
design
decisions.

\begin{figure}
\centering
\includegraphics[width=1.0\columnwidth]{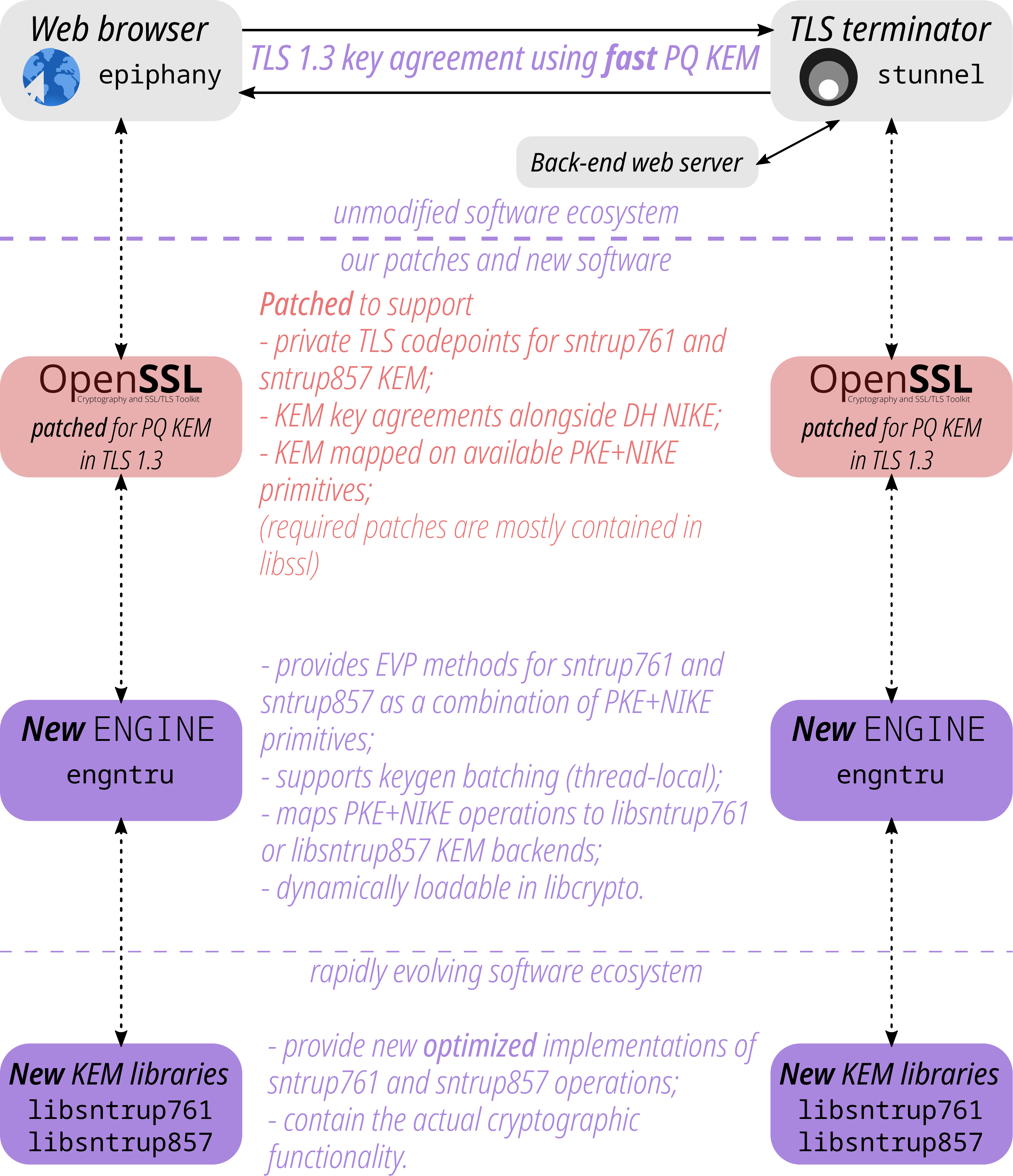}
\caption{Overview of our end-to-end experiment.}\label{fig:demo/overview}
\end{figure}

\autoref{fig:demo/overview} depicts a high-level overview of our
end-to-end
experiment,
highlighting
the
boundary
between
the
unmodified
software
ecosystem
and
our
novel
contributions.
This
section
details,
in
particular,
our
\openssl{}
patches
and
our
new
\ENGINE{} component.
\libsntrup{761} and \libsntrup{857} provide the new optimized
implementations
of
\sntrup[761]
and
\sntrup[857]
operations
(\autoref{sec:methods}),
through
a
simple
standardized
API
that
is
independent
from
\openssl{}, and reusable by other cryptographic software components.

\subsection{OpenSSL patches}\label{sec:ossl_patches}

\begin{figure}[t]
\centering
\includegraphics[width=1.0\columnwidth]{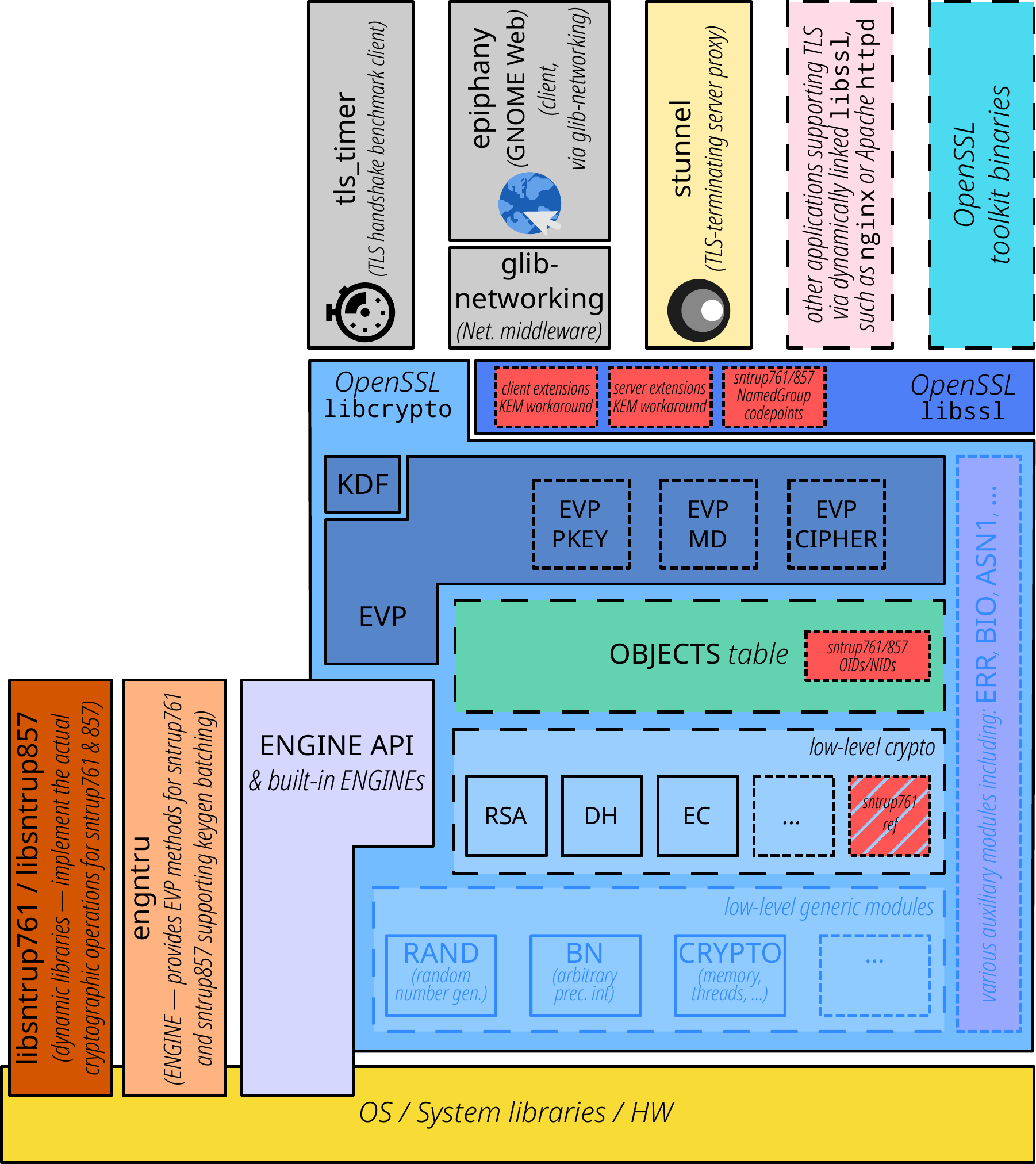}
\caption{Architecture diagram of the end-to-end experiment,
derived
from~\cite[Figure
2]{DBLP:conf/secdev/TuveriB19}.
The
red
boxes
within
\libssl{}
and
\libcrypto{}
represent
patches
applied
to
\opensslLTS{}
to
enable
our
post-quantum
KEM
experiment
over
\tlsonethree{}.
The
striped
\code{sntrup761
ref}
box
represents
the
optional
patch
to
also
include
a
reference
implementation
for
\sntrup[761]
inside
\libcrypto{}.
When
loaded,
\engntru{}
overrides
it
with
the
optimized
implementation
from
\libsntrup{761}.
}\label{fig:demo/architecture}
\end{figure}

\autoref{fig:demo/architecture} depicts an architecture diagram of our
end-to-end
experiment,
highlighting
with
red
boxes
inside
the
\libcrypto{} and \libssl{} modules, the patched \openssl{} components.

\Paragraph{\libssl{} changes}Within
\libssl{},
conceptually
three
elements
need
to
be
changed:
\begin{itemize}
\item Modify the server-side handling of the \keyshare{} extension
in
an
outgoing
\msg{ServerHello}
to
conditionally
use
a
KEM
\Encap{} operation for KEM groups;
\item Modify the client-side handling of the \keyshare{} extension
in
an
incoming
\msg{ServerHello}
to
conditionally
use
a
KEM
\Decap{} operation for KEM groups;
\item Hardcode private \code{NamedGroup} \tlsonethree{} codepoints
to
negotiate
\sntrup[761]
or
\sntrup[857]
groups
for
key
exchange.
\end{itemize}
As
\opensslLTS{}
does
not
provide
an
abstraction
for
KEM
primitives,
we
implemented
the
first
two
changes
as
a
workaround,
to
which
we
refer
as
PKE+NIKE.
It
maps
the
KEM
operations
as
a
combination
of
\emph{public-key encryption} (PKE) and
\emph{non-interactive key exchange} (NIKE).\footnote
{Generally
speaking,
the
\EVP{}
API
supports
any
NIKE
algorithm.
But
historically,
DH
and
ECDH
have
been
the
only
implementations
included
in
\openssl{}
for
this
API.
Hence,
code
and
documentation
tend
to
refer
to
such
primitives
as
\emph{DH key exchange} or just \emph{key exchange} rather than NIKE.}
We
combine
the
use
of
\evpencrypt{}
with
\NULL{}
input,
followed
by
\evpderive{} to mimic \Encap{}, and \evpdecrypt{} with \NULL{} output,
followed
by
\evpderive{}
for
\Decap{}.
Due
to
the
structure
of
the
PKE+NIKE
workaround,
on
both
sides
of
the
handshake,
handling
the
\keyshare{}
extension
for
KEM
groups
finishes
with
the
call
for
\evpderive{},
before
updating
the
protocol
key
schedule.
This
is
also
the
case
in
the
original
code
that
supports
traditional
NIKE.
Therefore,
the
new
code
only
affects
the
handling
of
the
opaque
\keyshare{} content transmitted over the wire.

As
depicted
in
\autoref{fig:tls13diagram},
on
the
server
side,
traditional
NIKE
groups
generate
an
ephemeral
key
pair,
sending
the
encoded
public
key
as
the
payload
of
the
extension.
With
our
patch,
if
the
group
is
flagged
as
a
KEM
group,
instead
of
key
generation
we
execute
\evpencrypt{}
under
the
client's
public
key
with
\NULL{}
input.
We
then
send
the
resulting
ciphertext
as
the
payload
of
the
\keyshare{}
extension.
As
a
side
effect,
per
our
PKE+NIKE
workaround,
\evpencrypt{}
also
stores
the
shared
secret
plaintext
within
the
internal
state
of
the
server-side
object
representing
the
client's
public
key.
This
plaintext
is
what
is
ultimately
retrieved
upon
calling
\evpderive{}.

On
the
client
side,
for
traditional
NIKE
groups,
the
payload
of
the
\keyshare{} extensions is parsed as the encoding of the peer's public
key,
to
be
used
in
the
subsequent
\evpderive{}.
With
our
patch,
if
the
group
is
flagged
as
a
KEM
group,
instead
we
treat
the
\keyshare{}
payload
as
the
ciphertext
to
be
used
in
\evpdecrypt{}
under
the
client's
secret
key,
and
with
\NULL{}
output.
The
resulting
plaintext
is
stored
in
the
internal
state
of
the
client-side
object
representing
the
client's
key
pair.
The
plaintext
shared
secret
is
ultimately
retrieved
via
\evpderive{}.

The
last
patch
alters
the
\libssl{}
static
table
of
supported
\tlsonethree{}
groups.
It
assigns
private
\code{NamedGroup}
codepoints
to
negotiate
\sntrup[761] or \sntrup[857] key exchanges, flagged as KEM groups,
and
links
it
to
static
numeric
identifiers
(\NID{}s)
defined
within
\libcrypto{}
headers.
These
identify
implementations
of
\sntrup[761] and \sntrup[857], as described in the next paragraph.

\Paragraph{\libcrypto{} changes}\libcrypto[\opensslLTSver] has the ability to generate
\NID{}s dynamically for custom algorithms unknown at \openssl{}
build
time.
In
contrast,
\libssl[\opensslLTSver]
defines
supported
groups
in
a
static
table
generated
during
compilation.
It
is
technically
possible
to
inject
KEM
functionality
(using
the
PKE+NIKE
workaround
described
above)
via
a
custom
\ENGINE{}
without
any
change
to
\libcrypto{}.
Yet,
the
limited
support
for
dynamic
customization
in
\libssl{}
adds
the
requirement
for
a
\libcrypto{} patch to issue static \NID{}s for \sntrup[761] and
\sntrup[857]. This is so they can be included in the \libssl{} static table
at
compile
time.
For
each
parameter
set,
this
patch
uses
the
internal
\openssl{}
tooling
to
issue
a
novel
static
\NID{} and associate it with the corresponding \sntrup[*] algorithm and a custom
\emph{object identifier} (\OID{}),\footnote
{\url{https://www.itu.int/en/ITU-T/asn1/Pages/OID-project.aspx}}
required
for
serializing
and
deserializing
key
objects.
With
this
data,
the
tooling
updates
the
public
\libcrypto{}
headers,
adding
the
relevant
\sntrup[*]
definitions.

Additionally,
we
include
an
optional
patch
for
\libcrypto{}
that
adds
a
reference
implementation
of
\sntrup[761]
as
a
new
\libcrypto{}
submodule.
Including
this
patch
allows
us
to
test
the
implementation
provided
by
\engntru{} against the reference implementation, and also to test the
software
stack
on
the
server
and
the
client
in
absence
of
the
\ENGINE{}.
This
eases
the
debug
process
during
the
development
of
\engntru{}.
For
the
final
users
of
our
end-to-end
scenario,
this
patch
is
entirely
optional,
as
the
dynamic
\ENGINE{}
injects
the
optimized
implementation
for
the
cryptographic
primitive
if
it
is
absent.
Otherwise,
it
overrides
the
default
reference
implementation
if
it
is
already
included
in
\libcrypto{}.

\subsection{The \engntru{} \ENGINE{}}\label{sec:engntru}

As
mentioned
in
\autoref{sec:related_integrations}
and
depicted
in
\autoref{fig:demo/overview} and \autoref{fig:demo/architecture}, as part
of
our
end-to-end
experiment,
we
introduce
a
new
\ENGINE{},
dubbed
\engntru{}.

We
followed
the
methodology
suggested
in
\cite{DBLP:conf/secdev/TuveriB19}, and we defer to it
for
a
detailed
description
of
the
\ENGINE{}
framework,
how
it
integrates
with
the
\openssl{}
architecture
(partially
illustrated
in
\autoref{fig:demo/architecture}),
security
considerations,
and
general
motivations
to
use
the
\ENGINE{}
framework
for
applied
research.
In
this
section,
we
highlight
how
this
choice
has
two
main
benefits:
it
decouples
\openssl{}
from
fast-paced
development
in
the
ecosystem
of
optimized
implementations
for
post-quantum
primitives,
and
at
the
same
time
it
decouples
external
libraries
implementing
novel
primitives
from
the
data
types
and
patterns
required
for
\openssl{}
compatibility.

\engntru{} builds upon \code{libbecc}~\cite{DBLP:conf/latincrypt/BrumleyHSTV19},
which
is
itself
derived
from
\code{libsuola}~\cite{DBLP:conf/secdev/TuveriB19}.
Similar
to
both
previous
works,
\engntru{} is also a \emph{shallow} \ENGINE{}, i.e., it does not contain
actual
cryptographic
implementations
for
the
supported
primitives.
Instead,
it
delegates
actual
computations
to
\libsntrup{761}
and
\libsntrup{857}.
The
functionality
provided
by
\engntru{}
includes:
\begin{itemize}
\item building as a dynamically loadable module, injecting support
for
novel
cryptographic
primitives
transparently
for
existing
applications;
\item supporting generic KEM primitives under the PKE+NIKE
workaround;
\item dynamically injecting/replacing support for \sntrup[761] at
run-time,
delegating
to
\libsntrup{761}
for
optimized
computation;
\item dynamically injecting support for \sntrup[857] at
run-time,
delegating
to
\libsntrup{857}
for
optimized
computation;
\item mapping the PKE+NIKE workaround back to the standard KEM API
adopted
by
the
implementations
of
NIST
PQC
KEM
candidates,
including
\libsntrup{*}.
\end{itemize}

Furthermore,
similar
to
\code{libbecc}
and
\code{libsuola},
and
using
the
same
terminology,
\engntru{}
supports
the
notion
of
multiple
providers
to
interface
with
the
\openssl{}
API.
Under
the
\code{serial\_lib}
provider,
each
\Keygen{}
operation
is
mapped
to
\function{crypto\_kem\_keypair},
generating
a
new
key
pair
on
demand
as
defined
by
the
NIST
PQC
KEM
API.
Alternatively,
under
the
\code{batch\_lib}
provider
(which
is
the
default
in
our
experiment),
\engntru{}
supports
batch
key
generation,
similar
to
\code{libbecc}.
In
the
case
of
\libsntrup{761}
and
\libsntrup{857}, this allows \openssl{} and applications to transparently
take
advantage
of
the
performance
gains
described
in
\autoref{sec:methods}.

Under
the
\code{batch\_lib}
model,
while
a
process
is
running,
each
\sntrup[*] parameter set is associated with a thread-safe
heap-allocated
pool
of
key
pairs.
Every
time
an
application
thread
requests
a
new
\sntrup[*] key pair, \engntru{} attempts to retrieve a fresh one from
the
corresponding
pool.
For
each
supported
parameter
set,
it
dynamically
allocates
a
pool,
initialized
the
first
time
a
key
pair
is
requested.
This
includes
filling
the
pool,
by
calling
\function{crypto\_kem\_sntrup761\_keypair\_batch} or
\function{crypto\_kem\_sntrup857\_keypair\_batch}.
Otherwise,
after
the
first
request,
\engntru{}
serves
each
request
by
copying
(and
then
securely
erasing
from
the
pool
buffer)
the
next
fresh
entry
in
the
pool.
After
this,
if
consuming
the
key
pair
emptied
the
pool,
\engntru{} fills it again, by calling the corresponding \libsntrup{*}
batch
generation
function.
This
happens
synchronously,
before
returning
control
to
the
application.
Storing
keys
for
deferred
use
adds
security
concerns:
\engntru{} addresses them relying on
standard
OS
guarantees
for
the
protection
of
memory
contents
across
processes
and
users.
On
the
other
hand,
the
batch
strategy
decouples
the
generation
of
a
key
pair
from
its
use
in
the
application
(e.g.,
an
attacker's
connection
request),
which
complicates
many
implementation
attacks,
and
results
in
an
overall
positive
security
impact.

In
terms
of
performance,
it
is
easy
to
see
the
advantage
of
\code{batch\_lib}
over
\code{serial\_lib}
from
our
microbenchmarks
in~\autoref{sec:methods}.
With
\code{serial\_lib},
each
\sntrup[761]
key
costs
$0.4$ms
on
a
2GHz
Haswell
core.
With
\code{batch\_lib},
within
each
batch
of
$32$
\sntrup[761]
keys,
the
first
key
costs
$2.5$ms,
and
the
remaining
$31$
keys
each
cost
$0$ms.
Note
that,
according
to
video-game
designers~\cite{2013/carmack},
latencies
below
20ms
are
imperceptible.
A
series
of
$K$
\sntrup[761]
keys
costs
$0.4K$ms
from~\code{serial\_lib}
and
just
$(0.08K+2.5)$ms
from~\code{batch\_lib}.
Similar
comments
apply
to
the
separate
\sntrup[857]
pool.

As
long
as
API/ABI
compatibility
is
maintained
in
the
\engntru{}/\libsntrup{*} interfaces, further refinements in the
\libsntrup{*} implementations do not require recompiling and reinstalling
\engntru{}, nor \openssl{}, nor other components of the software ecosystem
above.
At
the
same
time,
\libsntrup{761}
and
\libsntrup{857}
are
isolated
from
\openssl{}-specific
APIs,
so
they
can
easily
be
reused
by
alternative
stacks
supporting
the
NIST
PQC
KEM
API.
Moreover,
they
can
retain
a
lean
and
portable
API,
while
details
like
the
handling
of
pools
of
batch
results,
or
the
sharing
model
to
adopt,
are
delegated
to
the
middleware
layer.

\subsection{Reaching applications transparently}\label{sec:e2e:reachapps}

Consulting
\autoref{fig:demo/architecture},
the
purpose
of
this
section
is
to
describe
the
extent
of
the
application
layer
we
explored
in
our
study.
In
these
experiments,
we
investigated
two
paths
to
reach
\libssl{}
and
\libcrypto{}
(and
subsequently
\engntru{}
then
\libsntrup{*}).
Namely,
a
networking
application
dynamically
linking
directly,
and
a
separate
shared
library
against
which
even
higher
level
applications
dynamically
link
against.
More
generally,
this
approach
works
for
any
application
which
supports
\tlsonethree{} by dynamically linking against \libssl[1.1.1], but not
for
statically
linked
applications.\footnote
{
Although
not
part
of
our
end-to-end
demo
described
here,
we
further
validated
this
by
successfully
enabling
\sntrup{}
connections
in
popular
web
servers,
such
as
\code{nginx}
and
Apache
\code{httpd},
and
other
applications,
without
changes
to
their
sources
or
their
binary
distributions.
}

\Paragraph{stunnel}
For
networking
applications
that
do
not
natively
support
TLS,
\stunnel{}
is
an
application
that
provides
TLS
tunneling.
The
two
most
common
deployment
scenarios
for
\stunnel{}
are
client
mode
and
server
mode.

In
client
mode,
\stunnel{}
listens
for
cleartext
TCP
connections,
then
initiates
a
TLS
session
to
a
fixed
server
address.
A
common
use
case
for
client
mode
would
be
connecting
to
a
fixed
TLS
service
from
a
client
application
that
does
not
support
TLS.
For
example,
a
user
could
execute
the
\code{telnet}
application
(with
no
TLS
support)
to
connect
to
a
client
mode
instance
of
\stunnel{},
which
would
then
TLS-wrap
the
connection
to
a
static
SMTPS
server
to
securely
transfer
email.

In
server
mode,
\stunnel{}
listens
for
TLS
connections,
then
initiates
cleartext
TCP
connections
to
a
fixed
server
address.
A
common
use
case
for
server
mode
would
be
providing
a
TLS
service
from
a
server
application
that
does
not
support
TLS.
For
example,
a
user
could
serve
a
single
static
web
page
over
HTTP
with
the
\code{netcat} utility, which \stunnel{} would then TLS-wrap to serve the content
via
HTTPS
to
incoming
connections
from
e.g.\ browsers.
In
this
light,
\stunnel{}
server
mode
is
one
form
of
TLS
termination.

\stunnel{} links directly to \openssl{} for TLS functionality, hence the
intersection
with
\engntru{}
and
underlying
\libsntrup{*}
is
immediate.
For
example,
in
\stunnel{}
server
mode,
this
requires
no
changes
to
the
server
application,
which
in
fact
is
oblivious
to
the
TLS
tunneling
altogether.

\Paragraph{glib-networking}
\label{sec:e2e:middleware}
Similar
to
how
the
Standard
Template
Library
(STL)
and
Boost
provide
expanded
functionality
for
C++
(e.g.\ data
structures,
multithreading),
Glib
is
a
core
C
library
for
GNOME
and
GTK
applications.
Bundled
as
part
of
Glib,
one
feature
of
the
Gnome
Input/Output
(GIO)
C
library
provides
an
API
for
networking
functionality,
including
low-level
BSD-style
sockets.
For
TLS
connections,
GIO
loads
the
\GLN{}
C
library,
which
abstracts
away
the
backend
TLS
provider,
and
presents
a
unified
interface
to
callers.
Currently,
\GLN{}
supports
two
such
backends:
GnuTLS
and
\openssl{}.
The
latter
is
newer,
mainlined
in
\Ver{v2.59.90} (Feb 2019) while the current version as of this writing
is
\Ver{v2.68.1}.
This
is
precisely
the
place
where
\GLN{}
intersects
\openssl{}.
To
summarize,
the
modularity
of
\GLN{}
regarding
TLS
backends,
coupled
with
the
layered
approach
of
GIO,
allows
\emph{any}
application
utilizing
\GLN{}
for
TLS
functionality
to
\emph{transparently}
benefit
from
\ENGINE{}
features,
including
\engntru{}.

One
such
application,
and
one
highlight
of
our
experiment,
is
GNOME
Web.
Neither
Google
Chrome
nor
Mozilla
Firefox
are
capable
of
this
level
of
modularity.
Both
browsers
link
directly
to
TLS
backends
at
build
time
(BoringSSL,
NSS).
These
do
not
support
dynamically
injecting
this
level
of
cryptosystem
functionality,
necessarily
extending
to
the
TLS
layer
as
well.
In
general,
all
other
popular
browser
implementations
(we
are
aware
of)
require
source-code
changes
to
add
any
new
TLS
cipher
suite.
In
our
experiments,
we
are
able
to
make
GNOME
Web
\sntrup[761]-
and
\sntrup[857]-aware with
absolutely
no
changes
to
its
source
code,
nor
that
of
\GLN{}.
Performance-wise,
GNOME
Web
then
transparently
benefits
from
the
batch
key
generation
in
\libsntrup{*} through \engntru{}, loaded dynamically by the \openssl{} TLS
backend
of
\GLN{}.

\subsection{Macrobenchmarks: TLS handshakes}\label{sec:e2e_experiment:benchmark}

\begin{figure}
\centering
\includegraphics[width=1.0\columnwidth]{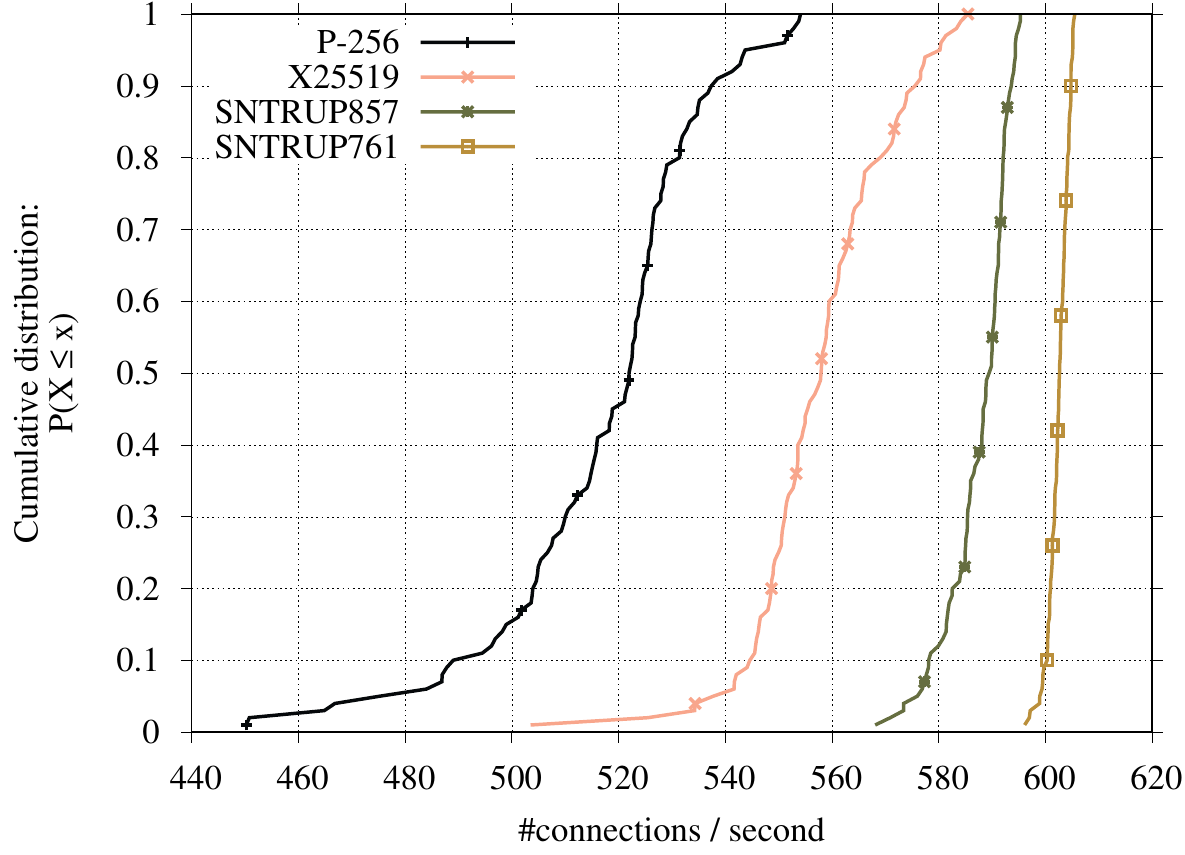}
\caption{Cumulative distributions of handshake performance under
different
cryptosystems
in
a
local
network.
Each
curve
represents
a
key-exchange
group,
for
which
we
collected
100
samples,
in
terms
of
average
number
of
connections
per
second.
This
metric
is
extrapolated
from
measuring
the
elapsed
wall-clock
time
over
8192
sequentially
established
connections
per
sample.
}\label{fig:e2e/measures}
\end{figure}

To
conclude
our
end-to-end
experiment,
we
investigated
the
impact
of
enabling
post-quantum
key
exchanges
for
\tlsonethree{}
handshakes,
as
perceived
by
end
users.
We
considered
an
experiment
on
large-scale
deployments
like
CECPQ1
or
CECPQ2
out
of
scope
for
this
work,
as
it
would
be
better
served
by
a
dedicated
study.
As
an
alternative,
we
decided
to
evaluate
the
performance
on
a
smaller
and
more
controlled
environment:
namely,
a
client
and
a
server
connected
over
a
low-traffic
Gigabit
Ethernet
network.
We
chose
to
focus
on
number
of
connections
per
second
as
the
more
relevant
metric
from
the
point
of
view
of
end
users,
and
used
easily
accessible
consumer
hardware
as
the
platform,
to
simulate
a
small
office
setup.\footnote
{
The
client
side
is
hosted
on
an
Intel
Core
i7-6700
workstation,
running
Ubuntu
20.04.2
with
Linux
5.4.0,
while
the
server
side
is
hosted
on
an
AMD
Ryzen
7
2700X
workstation,
running
Ubuntu
18.04.5
with
Linux
5.4.0.
Both
peers
directly
connect
to
the
same
Gigabit
Ethernet
L2
switch
via
their
embedded
Gigabit
Ethernet
NICs.
\label{footnote:e2e:hw}
}

To
exercise
full
control
over
the
sampling
process,
we
developed
a
small
(about
300
LOC)
TLS
client
built
directly
on
top
of
\libssl{} (see \autoref{apx:benchmarks} for a discussion about
in-browser
benchmarks).
Referring
to
the
diagram
in
\autoref{fig:demo/overview},
the
end-to-end
benchmark
replaces
\code{epiphany}
with
this
new
program,
that
we
dubbed
\code{tls\_timer}.
In
its
main
loop,
\code{tls\_timer}
records
a
timestamp,
sequentially
performs
a
predetermined
number
of
TLS
connections,
then
records
a
second
timestamp,
returning
the
elapsed
wall-clock
time.
In
the
above
loop,
for
each
connection,
it
performs
a
full
\tlsonethree{}
handshake.
Then,
the
client
properly
shuts
down
the
connection,
without
sending
any
application
data.
Hence,
the
total
elapsed
time
measured
by
the
client
covers
the
computation
time
required
by
client
and
server
to
generate
and
parse
the
content
of
the
exchanged
messages.
It
also
includes
the
time
spent
due
to
transit
of
packets
over
the
network,
and
through
userland/kernelspace
transitions.
In
particular,
with
respect
to
cryptographic
computations,
during
the
benchmark
the
client
repeatedly
performs
\Keygen{}
and
\Decap{}
for
the
ephemeral
key
exchange,
and
RSA-2048
signature
verifications
to
validate
the
identity
of
the
server
against
its
certificate.
During
the
client-measured
interval,
the
server
respectively
performs
\Encap{}
for
the
ephemeral
key
exchange,
and
RSA-2048
signature
generation
for
authentication.

As
a
baseline
for
comparisons,
we
used
\code{tls\_timer}
to
analogously
measure
the
performance
of
TLS
handshakes
using
the
most
popular
\tlsonethree{} groups for key exchange: namely,
\code{X25519} and \code{P-256}, in their respective ASM-optimized
implementations.
These
are
the
fastest
software
implementations
of
\tlsonethree{} key-exchange groups shipped in \openssl[1.1.1k], and are
widely
deployed
in
production.
For
these
groups,
computation
on
the
client
and
server
differs
from
the
description
above
exclusively
on
the
ephemeral
key
exchange,
as
both
sides
perform
their
respective
NIKE
\Keygen{}
and
\Derive{}
operations
instead
of
the
listed
post-quantum
KEM
operations,
as
summarized
in
\autoref{fig:tls13diagram}.

On
the
server
side
\code{tls\_timer}
connects
to
an
instance
of
\stunnel{}, configured as described above. Technically \stunnel{}
is
itself
connected
to
an
\code{apache2}
HTTP
daemon
serving
static
content
on
the
same
host,
but
as
\code{tls\_server}
does
not
send
any
application
data,
the
connection
between
\stunnel{}
and
\code{apache2}
is
short-lived
and
does
not
carry
data.
Finally,
to
minimize
noise
in
the
measurements,
we
disabled
frequency
scaling
and
Turbo
Boost
on
both
platforms,
terminated
most
concurrent
services
and
processes
on
the
client
and
the
server,
and
isolated
one
physical
core
exclusively
to
each
benchmark
process
(i.e.,
\code{tls\_timer}, \stunnel{} and \code{apache2}) to avoid biases due to
CPU
contention.

\autoref{fig:e2e/measures} visualizes our experimental results as
cumulative
distributions
for
each
tested
group.
The
results
show
that,
in
our
implementation,
both
the
recommended
\sntrup[761] parameter set and the higher security \sntrup[857]
consistently
achieve
more
connections
per
second
than
the
optimized
implementations
of
pre-quantum
alternatives
currently
deployed
at
large.

One
should
not
conclude
that
\sntrup[761]
and
\sntrup[857]
cost
less
than
ECC
overall.
The
unloaded
high-bandwidth
network
of
our
experimental
environment
masks
the
higher
communication
costs
of
the
lattice
cryptosystems,
whereas
reasonable
estimates
of
communication
costs
(see
generally~\autoref{hrss})
say
that
every
lattice
system
costs
more
than
ECC.
Nevertheless,
our
results
show
that,
in
terms
of
computational
costs,
we
achieve
new
records
when
compared
with
the
fastest
implementations
of
\tlsonethree{}
key-exchange
groups
included
in
\openssl[1.1.1k], while providing
higher
pre-quantum
security
levels
and
much
higher
post-quantum
security
levels
against
all
known
attacks.
This
significantly
reduces
the
total
\sntrup[*]
costs,
in
effect
assigning
higher
decision-making
weight
to
size
and,
most
importantly,
security.

\section{Conclusion} \label{sec:conclusion}

NIST's
ongoing
Post-Quantum
Cryptography
Standardization
Project
poses
significant
challenges
to
the
cryptology,
applied
cryptography,
and
system
security
research
communities,
to
name
a
few.
These
challenges
span
both
the
academic
and
industry
arenas.
Our
work
contributes
to
solving
these
challenges
in
two
main
directions.
(1)
In
\autoref{sec:methods},
we
propose
software
optimizations
for
\sntrup{},
from
fast
SIMD
arithmetic
at
the
lowest
level
to
efficient
amortized
batch
key
generation
at
the
highest
level.
These
are
an
essential
part
of
our
new
\libsntrup{761}
and
\libsntrup{857}
libraries.
(2)
In
\autoref{sec:e2e_experiment},
we
demonstrate
how
to
realize
these
gains
from
\libsntrup{*}
by
developing
\engntru{},
a
dynamically-loadable
\openssl{}
\ENGINE{}.
We
transparently
expose
it
to
the
application
layer
through
a
light
fork
of
\openssl{},
augmented
with
\sntrup{}
support
in
\tlsonethree{}
cipher
suites.
Our
experiments
reach
the
Gnome
Web
(\texttt{epiphany})
browser
on
the
client
side
and
\stunnel{}
as
a
TLS
terminator
on
the
server
side---both
with
no
source-code
changes.
Finally,
our
end-to-end
macrobenchmarks
combine
(1)
and
(2)
to
achieve
more
\tlsonethree{} handshakes per second than any software included in \openssl{}.

CECPQ1
and
CECPQ2
were
important
proof-of-concept
experiments
regarding
the
integration
of
post-quantum
algorithms
into
selected
browser
and
TLS
implementations,
but
those
experiments
suffered
from
poor
reproducibility:
the
capabilities
and
telemetry
are
only
available
to
major
industry
players
like
Google
and
Cloudflare,
so
the
cryptographic
primitive
choice
and
optimization
techniques
were
dictated
by
them
as
well.
Our
work
demonstrates
that
establishing
a
research
environment
to
provide
reproducible
results
is
not
only
feasible,
but
achievable
with
a
reasonable
workload
distribution,
using
new
TLS
software-layering
techniques
to
minimize
complexity
at
the
architecture
and
system
levels.

\Paragraph{Availability}
In
support
of
Open
Science,
we
provide
several
free
and
open-source
software
(FOSS)
contributions
and
research
artifacts.\relax
\footurl{https://opensslntru.cr.yp.to} \relax
We
released
\libsntrup{761},
\libsntrup{857},
\engntru{}, and \code{tls\_timer} as FOSS.
We
also
contributed
our
FOSS
implementations
of
enc
and
dec
to
SUPERCOP;
its
API
does
not
support
batch
keygen
at
this
time.
Lastly,
we
published
our
\openssl{}
patches
and
a
detailed,
step-by-step
tutorial
to
reproduce
our
full
experiment
stack.

\relax
\Paragraph{Acknowledgments}
This
work
was
funded
by
the
Deutsche
Forschungsgemeinschaft
(DFG,
German
Research
Foundation)
as
part
of
the
Excellence
Strategy
of
the
German
Federal
and
State
Governments---EXC
2092
CASA---390781972
``Cyber
Security
in
the
Age
of
Large-Scale
Adversaries'';
by
the
U.S.
National
Science
Foundation
under
grant
1913167;
by
the
Cisco
University
Research
Program;
and
by
the
European
Research
Council
(ERC)
under
the
European
Union's
Horizon
2020
research
and
innovation
programme
(grant
agreement
No
804476).
``Any
opinions,
findings,
and
conclusions
or
recommendations
expressed
in
this
material
are
those
of
the
author(s)
and
do
not
necessarily
reflect
the
views
of
the
National
Science
Foundation''
(or
other
funding
agencies).

The
\code{tls\_timer}
icon
used
in
\autoref{fig:demo/architecture}
is
licensed
under
CC-BY
and
created
by
Tomas
Knopp
for
\href{https://thenounproject.com}{thenounproject.com}.
All
product
names,
logos,
brands
and
trademarks
are
property
of
their
respective
owners.

The
scientific
colour
map
\textit{batlow}~\cite{colour:Crameri:SCM:7.0.0} is
used
in
this
study
to
prevent
visual
distortion
of
the
data
and
exclusion
of
readers
with
colour-vision
deficiencies
\cite{colour:Crameri:2020}.

\Paragraph{Metadata}
Date:
2021.10.06.
Permanent
ID
of
this
document:
{\tt
a8f6fc35a5dc11da1f125b9f5225d2a9c4c5b08b}.
\relax

\relax

\relax

\relax

\relax
\bibliographystyle{plnnt} \relax

\begin{thebibliography}{37}
\providecommand{\natexlab}[1]{#1}
\providecommand{\url}[1]{\texttt{#1}}
\expandafter\ifx\csname urlstyle\endcsname\relax
\providecommand{\doi}[1]{doi: #1}\else
\providecommand{\doi}{doi: \begingroup \urlstyle{rm}\Url}\fi

\bibitem[Alagic et~al.(2020)Alagic, Alperin-Sheriff, Apon, Cooper, Dang, Liu,
Miller,
Moody,
Peralta,
Perlner,
Robinson,
and
Smith-Tone]{2020/nist-8309}
G.~Alagic,
J.~Alperin-Sheriff,
D.~Apon,
D.~Cooper,
Q.~Dang,
Y.-K.
Liu,
C.~Miller,
D.~Moody,
R.~Peralta,
R.~Perlner,
A.~Robinson,
and
D.~Smith-Tone.
\newblock Status report on the second round of the {NIST Post-Quantum
Cryptography
Standardization
Process},
2020.
\newblock \url{https://csrc.nist.gov/publications/detail/nistir/8309/final}.

\bibitem[Alkim et~al.(2016)Alkim, Ducas, P{\"{o}}ppelmann, and
Schwabe]{2015/alkim}
E.~Alkim,
L.~Ducas,
T.~P{\"{o}}ppelmann,
and
P.~Schwabe.
\newblock
\href{https://www.usenix.org/conference/usenixsecurity16/technical-sessions/presentation/alkim}{Post-quantum
key
exchange
-
{A}
new
hope}.
\newblock In \emph{USENIX Security 2016}, pages 327--343, 2016.

\bibitem[Alkim et~al.(2021)Alkim, Cheng, Chung, Evkan, Huang, Hwang, Li,
Niederhagen,
Shih,
Wälde,
and
Yang]{2020/alkim-polymult}
E.~Alkim,
D.~Y.-L.
Cheng,
C.-M.~M.
Chung,
H.~Evkan,
L.~W.-L.
Huang,
V.~Hwang,
C.-L.~T.
Li,
R.~Niederhagen,
C.-J.
Shih,
J.~Wälde,
and
B.-Y.
Yang.
\newblock \href{https://doi.org/10.46586/tches.v2021.i1.217-238}{Polynomial
multiplication
in
{NTRU
Prime}:
Comparison
of
optimization
strategies
on
{Cortex-M4}}.
\newblock \emph{TCHES}, 2021\penalty0 (1):\penalty0 217--238, 2021.

\bibitem[Avanzi et~al.(2020)Avanzi, Bos, Ducas, Kiltz, Lepoint, Lyubashevsky,
Schanck,
Schwabe,
Seiler,
and
Stehl\'e]{2019/kyber}
R.~Avanzi,
J.~Bos,
L.~Ducas,
E.~Kiltz,
T.~Lepoint,
V.~Lyubashevsky,
J.~M.
Schanck,
P.~Schwabe,
G.~Seiler,
and
D.~Stehl\'e.
\newblock {CRYSTALS-Kyber: algorithm specifications and supporting
documentation},
2020.
\newblock
\url{https://csrc.nist.gov/projects/post-quantum-cryptography/round-3-submissions}.

\bibitem[Basso et~al.(2020)Basso, Mera, D'Anvers, Karmakar, Roy, Beirendonck,
and
Vercauteren]{2019/saber}
A.~Basso,
J.~M.~B.
Mera,
J.-P.
D'Anvers,
A.~Karmakar,
S.~S.
Roy,
M.~V.
Beirendonck,
and
F.~Vercauteren.
\newblock {SABER: Mod-LWR based KEM (round 3 submission)}, 2020.
\newblock
\url{https://csrc.nist.gov/projects/post-quantum-cryptography/round-3-submissions}.

\bibitem[Bellare et~al.(2020)Bellare, Davis, and G{\"{u}}nther]{2020/bellare}
M.~Bellare,
H.~Davis,
and
F.~G{\"{u}}nther.
\newblock \href{https://doi.org/10.1007/978-3-030-45724-2_1}{Separate your
domains:
{NIST}
{PQC}
{KEMs},
oracle
cloning
and
read-only
indifferentiability}.
\newblock In \emph{EUROCRYPT 2020}, pages 3--32, 2020.
\newblock \url{https://eprint.iacr.org/2020/241}.

\bibitem[Bernstein(2001)]{2001/bernstein}
D.~J.
Bernstein.
\newblock Multidigit multiplication for mathematicians, 2001.
\newblock \url{http://cr.yp.to/papers.html#m3}.

\bibitem[Bernstein(2009)]{djb09:karatsuba}
D.~J.
Bernstein.
\newblock \href{https://doi.org/10.1007/978-3-642-03356-8_19}{Batch binary
{Edwards}}.
\newblock In S.~Halevi, editor, \emph{CRYPTO 2009}, volume 5677 of
\emph{Lecture Notes in Computer Science}, pages 317--336. Springer, 2009.

\bibitem[Bernstein and Lange(2020)]{2020/bernstein-horror}
D.~J.
Bernstein
and
T.~Lange.
\newblock Crypto horror stories, 2020.
\newblock \url{https://hyperelliptic.org/tanja/vortraege/20200206-horror.pdf}.

\bibitem[Bernstein and Lange(2021)]{2021/bernstein-ebacs}
D.~J.
Bernstein
and
T.~Lange.
\newblock {eBACS}: {ECRYPT} benchmarking of cryptographic systems, 2021.
\newblock \url{https://bench.cr.yp.to}, accessed 28 May 2021.

\bibitem[Bernstein and Yang(2019)]{BY19:extgcd}
D.~J.
Bernstein
and
B.~Yang.
\newblock \href{https://doi.org/10.13154/tches.v2019.i3.340-398}{Fast
constant-time
gcd
computation
and
modular
inversion}.
\newblock \emph{TCHES}, 2019\penalty0 (3):\penalty0 340--398, 2019.

\bibitem[Bernstein et~al.(2020)Bernstein, Brumley, Chen, Chuengsatiansup,
Lange,
Marotzke,
Peng,
Tuveri,
van
Vredendaal,
and
Yang]{2020/ntruprime}
D.~J.
Bernstein,
B.~B.
Brumley,
M.-S.
Chen,
C.~Chuengsatiansup,
T.~Lange,
A.~Marotzke,
B.-Y.
Peng,
N.~Tuveri,
C.~van
Vredendaal,
and
B.-Y.
Yang.
\newblock {NTRU Prime: round 3}, 2020.
\newblock
\url{https://csrc.nist.gov/projects/post-quantum-cryptography/round-3-submissions}.

\bibitem[Biasse and Song(2016)]{2016/biasse}
J.~Biasse
and
F.~Song.
\newblock \href{http://dx.doi.org/10.1137/1.9781611974331.ch64}{Efficient
quantum
algorithms
for
computing
class
groups
and
solving
the
principal
ideal
problem
in
arbitrary
degree
number
fields}.
\newblock In \emph{SODA 2016}, pages 893--902, 2016.

\bibitem[Bindel et~al.(2019)Bindel, Brendel, Fischlin, Goncalves, and
Stebila]{DBLP:conf/pqcrypto/BindelBFGS19}
N.~Bindel,
J.~Brendel,
M.~Fischlin,
B.~Goncalves,
and
D.~Stebila.
\newblock \href{https://doi.org/10.1007/978-3-030-25510-7_12}{Hybrid key
encapsulation
mechanisms
and
authenticated
key
exchange}.
\newblock In \emph{PQCrypto 2019}, pages 206--226. Springer, 2019.

\bibitem[Brumley et~al.(2019)Brumley, ul~Hassan, Shaindlin, Tuveri, and
Vuoj{\"{a}}rvi]{DBLP:conf/latincrypt/BrumleyHSTV19}
B.~B.
Brumley,
S.~ul~Hassan,
A.~Shaindlin,
N.~Tuveri,
and
K.~Vuoj{\"{a}}rvi.
\newblock \href{https://doi.org/10.1007/978-3-030-30530-7_18}{Batch binary
{Weierstrass}}.
\newblock In \emph{{LATINCRYPT}}, pages 364--384. Springer, 2019.

\bibitem[Carmack(2013)]{2013/carmack}
J.~Carmack.
\newblock Latency mitigation strategies, 2013.
\newblock
\url{https://web.archive.org/web/20130225013015/www.altdevblogaday.com/2013/02/22/latency-mitigation-strategies/}.

\bibitem[Chen et~al.(2020)Chen, Danba, Hoffstein, Hulsing, Rijneveld, Schanck,
Saito,
Schwabe,
Whyte,
Yamakawa,
Xagawa,
and
Zhang]{2020/ntru}
C.~Chen,
O.~Danba,
J.~Hoffstein,
A.~Hulsing,
J.~Rijneveld,
J.~M.
Schanck,
T.~Saito,
P.~Schwabe,
W.~Whyte,
T.~Yamakawa,
K.~Xagawa,
and
Z.~Zhang.
\newblock {NTRU: algorithm specifications and supporting documentation}, 2020.
\newblock
\url{https://csrc.nist.gov/projects/post-quantum-cryptography/round-3-submissions}.

\bibitem[Chung et~al.(2021)Chung, Hwang, Kannwischer, Seiler, Shih, and
Yang]{2021/Chung}
C.~M.
Chung,
V.~Hwang,
M.~J.
Kannwischer,
G.~Seiler,
C.~Shih,
and
B.~Yang.
\newblock \href{https://doi.org/10.46586/tches.v2021.i2.159-188}{{NTT}
multiplication
for
{NTT}-unfriendly
rings:
New
speed
records
for
{Saber}
and
{NTRU}
on
{Cortex-M4}
and
{AVX2}}.
\newblock \emph{TCHES}, 2021\penalty0 (2):\penalty0 159--188, 2021.

\bibitem[Crameri(2021)]{colour:Crameri:SCM:7.0.0}
F.~Crameri.
\newblock Scientific colour maps, February 2021.
\newblock \url{https://doi.org/10.5281/zenodo.4491293}.

\bibitem[Crameri et~al.(2020)Crameri, Shephard, and Heron]{colour:Crameri:2020}
F.~Crameri,
G.~E.
Shephard,
and
P.~J.
Heron.
\newblock \href{https://doi.org/10.1038/s41467-020-19160-7}{The misuse of
colour
in
science
communication}.
\newblock \emph{Nature Communications}, 11\penalty0 (1):\penalty0 5444, Oct
2020.

\bibitem[Fog(2021)]{fog2021instruction}
A.~Fog.
\newblock \emph{Instruction tables: Lists of instruction latencies, throughputs
and
micro-operation
breakdowns
for
{Intel},
{AMD}
and
{VIA}
{CPUs}}.
\newblock Technical University of Denmark, March 2021.
\newblock \url{https://www.agner.org/optimize/instruction_tables.pdf}, accessed
22
March
2021.

\bibitem[Kannwischer et~al.(2021)Kannwischer, Rijneveld, Schwabe, Stebila, and
Wiggers]{github/pqclean}
M.~Kannwischer,
J.~Rijneveld,
P.~Schwabe,
D.~Stebila,
and
T.~Wiggers.
\newblock {PQClean}: clean, portable, tested implementations of postquantum
cryptography,
2021.
\newblock \url{https://github.com/pqclean/pqclean}.

\bibitem[Menezes et~al.(1996)Menezes, Vanstone, and Oorschot]{HAC96}
A.~J.
Menezes,
S.~A.
Vanstone,
and
P.~C.~V.
Oorschot.
\newblock \emph{Handbook of Applied Cryptography}.
\newblock CRC Press, Inc., USA, 1st edition, 1996.

\bibitem[Montgomery(1987)]{Montgomery1987SpeedingTP}
P.~Montgomery.
\newblock Speeding the {Pollard} and elliptic curve methods of factorization.
\newblock \emph{Math. Comp.}, 48:\penalty0 243--264, 1987.

\bibitem[Montgomery(1985)]{85/Montgomery}
P.~L.
Montgomery.
\newblock \href{https://doi.org/10.1090/S0025-5718-1985-0777282-X}{Modular
multiplication
without
trial
division}.
\newblock \emph{Math. Comp.}, 44:\penalty0 519--521, 1985.

\bibitem[NIST(2020)]{NIST:guideline}
NIST.
\newblock Guidelines for submitting tweaks for third round finalists and
candidates,
2020.
\newblock
\url{https://csrc.nist.gov/CSRC/media/Projects/post-quantum-cryptography/documents/round-3/guidelines-for-sumbitting-tweaks-third-round.pdf}.

\bibitem[{NIST PQC team}(2020)]{2020/nist-tweaks}
{NIST
PQC
team}.
\newblock Guidelines for submitting tweaks for {Third Round Finalists} and
{Candidates},
2020.
\newblock
\url{https://groups.google.com/a/list.nist.gov/g/pqc-forum/c/LPuZKGNyQJ0/m/O6UBanYbDAAJ}.

\bibitem[Nussbaumer(1980)]{80/Nussbaumer}
H.~Nussbaumer.
\newblock \href{https://doi.org/10.1109/TASSP.1980.1163372}{Fast polynomial
transform
algorithms
for
digital
convolution}.
\newblock \emph{IEEE ASSP}, 28:\penalty0 205--215, 1980.

\bibitem[Paquin et~al.(2020)Paquin, Stebila, and
Tamvada]{DBLP:conf/pqcrypto/PaquinST20}
C.~Paquin,
D.~Stebila,
and
G.~Tamvada.
\newblock \href{https://doi.org/10.1007/978-3-030-44223-1_5}{Benchmarking
post-quantum
cryptography
in
{TLS}}.
\newblock In \emph{PQCrypto}, pages 72--91. Springer, 2020.

\bibitem[P{\"{u}}schel et~al.(2005)P{\"{u}}schel, Moura, Johnson, Padua,
Veloso,
Singer,
Xiong,
Franchetti,
Gacic,
Voronenko,
Chen,
Johnson,
and
Rizzolo]{2005/spiral}
M.~P{\"{u}}schel,
J.~M.~F.
Moura,
J.~R.
Johnson,
D.~A.
Padua,
M.~M.
Veloso,
B.~Singer,
J.~Xiong,
F.~Franchetti,
A.~Gacic,
Y.~Voronenko,
K.~Chen,
R.~W.
Johnson,
and
N.~Rizzolo.
\newblock \href{https://doi.org/10.1109/JPROC.2004.840306}{{SPIRAL:} code
generation
for
{DSP}
transforms}.
\newblock \emph{Proc. {IEEE}}, 93\penalty0 (2):\penalty0 232--275, 2005.

\bibitem[Sch\"{o}nhage(1977)]{77/Schoennhage}
A.~Sch\"{o}nhage.
\newblock \href{https://doi.org/10.1007/BF00289470}{Schnelle {Multiplikation}
von
{Polynomen}
\"{u}ber
{K\"{o}rpern}
der
{Charakteristik}
2}.
\newblock \emph{Acta Inf.}, 7\penalty0 (4):\penalty0 395–398, December 1977.

\bibitem[Schwabe et~al.(2020)Schwabe, Stebila, and
Wiggers]{DBLP:conf/ccs/SchwabeSW20}
P.~Schwabe,
D.~Stebila,
and
T.~Wiggers.
\newblock \href{https://doi.org/10.1145/3372297.3423350}{Post-quantum {TLS}
without
handshake
signatures}.
\newblock In \emph{{ACM} {CCS}}, pages 1461--1480. {ACM}, 2020.

\bibitem[Shacham and Boneh(2001)]{SB02:batching}
H.~Shacham
and
D.~Boneh.
\newblock Improving {SSL} handshake performance via batching.
\newblock In \emph{CT-RSA 2001}, page 28–43, 2001.
\newblock \url{https://hovav.net/ucsd/papers/sb01.html}.

\bibitem[Stebila and Mosca(2016)]{DBLP:conf/sacrypt/StebilaM16}
D.~Stebila
and
M.~Mosca.
\newblock \href{https://doi.org/10.1007/978-3-319-69453-5_2}{Post-quantum key
exchange
for
the
{Internet}
and
the
{Open
Quantum
Safe}
project}.
\newblock In \emph{{SAC}}, pages 14--37. Springer, 2016.

\bibitem[Stebila et~al.(2021)Stebila, Fluhrer, and
Gueron]{ietf-tls-hybrid-design}
D.~Stebila,
S.~Fluhrer,
and
S.~Gueron.
\newblock Hybrid key exchange in {TLS} 1.3, April 2021.
\newblock
\url{https://datatracker.ietf.org/doc/html/draft-ietf-tls-hybrid-design-02}.

\bibitem[Tuveri and Brumley(2019)]{DBLP:conf/secdev/TuveriB19}
N.~Tuveri
and
B.~B.
Brumley.
\newblock \href{https://doi.org/10.1109/SecDev.2019.00014}{Start your
{ENGINEs}:
Dynamically
loadable
contemporary
crypto}.
\newblock In \emph{SecDev}, pages 4--19. {IEEE}, 2019.

\bibitem[Warren(2012)]{2012/hackersdelight}
H.~S.
Warren.
\newblock \emph{Hacker's Delight}.
\newblock Addison-Wesley Professional, 2nd edition, 2012.

\end{thebibliography}

\relax

\relax

\appendix
\section{Further improvements in NTRU Prime software}
\label{furthersoftware}

This
paper
emphasizes
a
big
speedup
in
{\tt
sntrup}
key
generation,
and
new
software
layers
integrating
this
speedup
into
TLS
software.
The
speedup
relies
on
changing
the
key-generation
API
to
generate
many
keys
at
once,
and
providing
one
key
at
a
time
on
top
of
this
requires
maintaining
state,
which
is
enabled
by
the
new
software
layers.

This
appendix
describes
other
ways
that
we
have
improved
the
NTRU
Prime
software
{\it
without\/}
changing
the
API.
The
software
was
already
heavily
tuned
before
our
work,
but
some
further
streamlining
turned
out
to
be
possible,
for
example
reducing
\sntrup[761]
enc
from
48892
cycles
to
{\ourenc}
cycles
and
reducing
dec
from
59404
cycles
to
{\ourdec}
cycles.
More
important
than
these
quantitative
speedups
is
the
software
engineering:
we
considerably
simplified
the
preexisting
optimized
code
base
for
NTRU
Prime,
especially
with
the
new
NTT
compiler
described
below.

\Paragraph{Review of NTRU Prime options}
The
NTRU
Prime
proposal
specifies
various
lattice
dimensions.
Round-1
NTRU
Prime
specified
only
dimension
761.
Round-2
NTRU
Prime
specified
dimensions
653,
761,
and
857.
Round-3
NTRU
Prime---which
appeared
after
our
first
announcement
of
the
OpenSSLNTRU
results---also
specified
953,
1013,
and
1277.

NTRU
Prime
specifies
two
cryptosystems:
Streamlined
NTRU
Prime
({\tt
sntrup}),
an
example
of
Quotient
NTRU,
and
NTRU
LPRime
({\tt
ntrulpr}),
an
example
of
Product
NTRU.
For
example,
dimension
761
has
both
{\tt
sntrup761}
and
{\tt
ntrulpr761}.
The
two
cryptosystems
are
almost
identical
in
key
sizes,
ciphertext
sizes,
and
Core-SVP
security.
The
{\tt
ntrulpr}
cryptosystems
avoid
the
Quotient
NTRU
inversions
and
have
much
faster
keygen
than
{\tt
sntrup},
but
they
have
slower
enc
and
slower
dec
than
{\tt
sntrup}.

\Paragraph{Preexisting AVX2-optimized software}
The
official
NTRU
Prime
``optimized
C
software''
uses
Intel
AVX2
instructions
and
supports
dimensions
653,
761,
857.
Some
of
the
code
is
shared
across
sizes
except
for
compile-time
selection
of
$q$
etc.
There
is
less
sharing
of
the
multiplier
code
across
sizes:
dimensions
653
and
761
use
{\tt
mult768.c},
which
uses
size-512
NTTs
to
multiply
768-coefficient
polynomials;
dimension
857
uses
{\tt
mult1024.c},
which
uses
size-512
NTTs
to
multiply
1024-coefficient
polynomials.
An
underlying
{\tt
ntt.c}
is
shared
for
computing
size-512
NTTs,
and
the
same
NTT
code
is
used
for
each
of
the
NTT-friendly
primes
$r\in\setof{7681,10753}$,
but
multiplication
algorithms
vary
between
{\tt
mult768.c}
and
{\tt
mult1024.c}:
for
example,
{\tt
mult768.c}
uses
``Good's
trick''
to
reduce
a
size-1536
NTT
to
$3$
size-512
NTTs,
taking
advantage
of
$3$
being
odd,
while
{\tt
mult1024.c}
uses
a
more
complicated
method
to
reduce
a
size-2048
NTT
to
$4$
size-512
NTTs.
The
NTT
API
allows
these
$3$
or
$4$
independent
size-512
NTTs
to
be
computed
with
one
function
call,
reducing
per-call
overheads
and
also
reducing
the
store-to-load-forwarding
overheads
in
crossing
NTT
layers.

\Paragraph{Improvements}
We
first
built
a
tool
to
regenerate
653,
761,
and
857
in
the
optimized
C
implementation
from
a
merged
code
base.
We
then
added
support
for
953,
1013,
and
1277,
which
in
previous
work
had
only
reference
code.
This
meant,
among
other
things,
building
a
new
{\tt
mult1280.c}
to
reduce
a
size-2560
NTT
to
$5$
size-512
NTTs.
Good's
trick
is
applicable
here
since
$5$
is
odd,
but
we
were
faced
with
a
new
mini-optimization
problem
regarding
the
number
of
AVX2
instructions
needed
for
$5$-coefficient
polynomial
multiplications
modulo
$r$.
The
best
solution
we
found
uses
$15$
modular
multiplications,
$2$
extra
reductions,
and
$34$
additions/subtractions.

We
then
built
a
new
tool
to
compile
concise
descriptions
of
NTT
strategies
into
optimized
NTT
software.
This
tool
is
analogous
to
SPIRAL~\cite{2005/spiral},
but
handles
the
extra
complications
of
NTTs
compared
to
floating-point
FFTs,
notably
the
requirement
of
tracking
ranges
of
intermediate
quantities
so
as
to
avoid
overflows.
Note
that
one
should
not
confuse
automated
generation
of
NTTs
with
automated
generation
of
multipliers;
it
remains
challenging
to
automate
code
generation
for
the
type
of
multipliers
that
we
consider
in
\autoref{sec:methods}.

Armed
with
this
tool,
we
searched
for
efficient
size-512
NTT
strategies
to
replace
the
previous
{\tt
ntt.c}.
We
found
a
fully
vectorizable
strategy
that
avoids
all
overflows
for
both
$r=7681$
and
$r=10753$;
uses
just
6656
16-bit
multiplications;
uses
just
6976
16-bit
additions
(counting
subtractions
as
additions);
stores
data
only
every
3
NTT
layers;
and
has
only
4
layers
of
permutation
instructions.
To
put
this
in
perspective,
if
each
of
the
$9$
NTT
layers
had
$256$
modular
multiplications,
$512$
additions,
and
zero
extra
modular
reductions,
then
in
total
there
would
be
6912
16-bit
multiplications
and
6912
16-bit
additions,
since
each
modular
multiplication
costs
3
16-bit
multiplications
and
1
16-bit
addition.

\section{Comparing {\tt ntruhrss}}
\label{hrss}

CECPQ2's
{\tt
ntruhrss701}
keygen,
like
OpenSSLNTRU's
{\tt
sntrup761}/{\tt
sntrup857}
keygen,
is
bottlenecked
by
inversion.
Conceptually,
everything
this
paper
does
for
\sntrup[]
can
also
be
done
for
\ntruhrss[],
starting
with
converting
a
batch
of
32
{\tt
ntruhrss}
inversions
into
1
inversion
plus
93
multiplications.
This
appendix
explains
two
factors
making
this
strategy
less
attractive
for
{\tt
ntruhrss}
compared
to
{\tt
sntrup}.

First,
a
reasonable
estimate,
based
on
a
close
look
at
the
underlying
algorithms,
is
that
there
would
be
only
about
a
$2\times$
speedup
from
\ntruhrss[]
keygen
to
batched
\ntruhrss[]
keygen,
much
less
than
the
speedup
we
achieve
for
\sntrup[].

The
reason
is
as
follows.
Unbatched
keygen
is
bottlenecked
by
inversion,
and
\ntruhrss[]
exploits
one
of
its
design
decisions---a
power-of-$2$
modulus,
which
\sntrup[]
avoids
because
of
security
concerns---for
a
specialized
type
of
inversion
algorithm,
a
``Hensel
lift''.
Batched
keygen
is
instead
bottlenecked
by
multiplication,
and
benefits
much
less
from
a
power-of-$2$
modulus.
The
Hensel
speedup
would
still
be
measurable
inside
the
occasional
inversion,
but
batch
size
$32$
compresses
this
speedup
by
a
factor
$32$.
One
can
see
{\it
some\/}
speedup
from
{\tt
sntrup761}
to
{\tt
ntruhrss701}
in
multiplications
(because
of
the
modulus
and
the
lower
\ntruhrss[701]
security
level),
but
the
ultimate
difference
in
keygen
speeds
will
be
an
order
of
magnitude
smaller
than
the
difference
in
keygen
speeds
before
this
paper.

Second,
the
network-traffic-vs.-security-level
trade-off
is
worse
for
\ntruhrss[]
than
for
\sntrup[].
For
example,
\autoref{comparison}
shows
that
{\tt
ntruhrss701}
sends
3.6\%
more
traffic
than
{\tt
sntrup761},
despite
having
only
89\%
of
the
security
level
(Core-SVP
$2^{136}$
vs.~$2^{153}$).

An
existing
cost
model
estimates
that
1000
CPU
cycles
have
the
same
cost
as
communicating
a
byte
of
data:
e.g.,
a
quad-core
3GHz
server
has
the
same
cost
as
a
100Mbps
Internet
connection.
An
easy
calculation
from~\autoref{comparison}
concludes
that
higher-security
\sntrup[761]
would
still
cost
1.8\%
below
\ntruhrss[701]
in
this
model
even
after
a
$2\times$
speedup
in
\ntruhrss[701]
keygen.
Making
\ntruhrss[]
competitive
with
\sntrup[],
accounting
for
the
security
level,
would
require
not
just
this
speedup
but
also
focusing
on
environments
where
communication
is
more
than
$10\times$
cheaper.

\section{Barrett reduction correctness}
\label{sec:proofmod3}

Recalling
\autoref{sec:iszero},
Barrett
reduction
estimates
$g/f_i$
as
$
h
=
\floor{
g/f_i
}
$.
Then
it
calculates
the
remainder
as
$
r
=
g
-
g
\cdot
\floor{
{g}/{f_i}
}
\cdot
f_i
$.
We
compute
the
difference
between
$({g}/{f_i})\cdot
f_i$
and
$
\floor{
{g}/{f_i}
}
\cdot
f_i$.
It
is
the
remainder
$r$
if
the
difference
has
degree
less
than
$\deg(f_i)$.

Using
the
pre-computation
$
x^{D_g}
=
q_x
\cdot
f_i
+
r_x
$,
we
have
\[
{g}/{f_i}
=
 g
\cdot
({
x^{D_g}
}/{f_i})
\cdot
{1}/{
x^{D_g}
}
=
g
\cdot
(
q_x
-
{r_x}/{f_i}
)
\cdot
{1}/{
x^{D_g}
}
.
\]
We
compute
the
difference
\relax
\begin{equation}
\label{eq:diff}
\begin{aligned}
&
({g}/{f_i})\cdot
f_i
-
 \floor{
{g}/{f_i}
}
\cdot
f_i
\\&\quad
=
(
{
g
\cdot
q_x
}/{x^{D_g}}
-
\floor{
 {
g
\cdot
q_x
}/{x^{D_g}}
}
)
\cdot
f_i
-
g
 ({r_x}/{f_i})
 {
f_i
}/{x^{D_g}}
.
\end{aligned}
\end{equation}
\relax

Define
$h
=
\floor{
{
g
\cdot
q_x
}/{x^{D_g}}
}
=
 \floor{
{g}/{f_i}
}
$
and
$
l
=
{
g
\cdot
q_x
}/{x^{D_g}}
-
h
$.
The
term
$
(
{
g
\cdot
q_x
 }/{x^{D_g}}
-
\floor{
 {
g
\cdot
q_x
}/{x^{D_g}}
}
)
\cdot
f_i
=
l
\cdot
f_i
 $
in
\autoref{eq:diff}
has
degree
less
than
$\deg(f_i)$.
The
other
term
$g
 ({r_x}/{f_i})
 {
f_i
}/{x^{D_g}}$
also
has
degree
less
than
$\deg(f_i)
$,
since
$
\deg(g)
<
D_g
$
and
$
\deg(r_x)
<
\deg(f_i)
$.

\section{More on benchmarks}\label{apx:benchmarks}

\Paragraph{Batch key-generation microbenchmarks}
\autoref{tab::keygen:performance:2} shows the performance and key pair storage
of
$\BatchKeyGen$
regarding
various
batch
sizes
$n$.

\begin{table*}[h]
\caption{Performance of $\BatchKeyGen$ regarding various batch sizes $n$.}\label{tab::keygen:performance:2}
\resizebox{1.0\textwidth}{!}{\begin{tabular}{ |c | c || r || r | r | r | r | r | r | r |  }
\hline
&
 $n$

&
1
 &
2
&
4
&
8
&
16
 &
32
 &
64
 &
128
\\
\hline
\hline
\multirow{4}{*}{\sntrup[653]}
&
amortized
cost$^\dag$
&
\num{778218}
&
\num{438714}
&
\num{295150}
&
\num{229429}
&
\num{180863}
&
\num{164260}
&
\num{152737}
&
\num{147821}
\\
\cline{2-10}
&
latency$^\dag$
&
\num{778218}
&
\num{877428}
&
\num{1180600}
&
\num{1835432}
&
\num{2893808}
&
\num{5256300}
&
\num{9775160}
&
\num{18921036}
\\
\cline{2-10}
&
key
pair
storage$^\ddag$
&
\num{2512}
&
\num{5024}
&
\num{10048}
&
\num{20096}
&
\num{40192}
&
\num{80384}
&
\num{160768}
&
\num{321536}
\\
\cline{2-10}
&
memory
footprint$^{\ddag\star}$
&
\num{30432}
&
\num{36184}
&
\num{54808}
&
\num{65432}
&
\num{85880}
&
\num{127672}
&
\num{211480}
&
\num{378424}
\\
\hline
\hline
\multirow{4}{*}{\sntrup[761]}
&
amortized
cost$^\dag$
&
\num{819332}
&
\num{567996}
&
\num{351329}
&
\num{242043}
&
\num{181274}
&
\num{
\ourkeygen
}
&
\num{147809}
&
\num{141411}
\\
\cline{2-10}
&
latency$^\dag$
&
\num{819332}
&
\num{1135992}
&
\num{1405316}
&
\num{1936340}
&
\num{2900380}
&
\num{5002124}
&
\num{9459748}
&
\num{18100592}
\\
\cline{2-10}
&
key
pair
storage$^\ddag$
&
\num{2921}
&
\num{5842}
&
\num{11684}
&
\num{23368}
&
\num{46736}
&
\num{93472}
&
\num{186944}
&
\num{373888}
\\
\cline{2-10}
&
memory
footprint$^{\ddag\star}$
&
\num{117200}$^\ast$
&
\num{38608}
&
\num{58040}
&
\num{70456}
&
\num{94584}
&
\num{143288}
&
\num{240536}
&
\num{435512}
\\
\hline
\hline
\multirow{4}{*}{\sntrup[857]}
&
amortized
cost$^\dag$
&
\num{1265056}
&
\num{708104}
&
\num{458562}
&
\num{322352}
&
\num{255815}
&
\num{216618}
&
\num{201173}
&
\num{193203}
\\
\cline{2-10}
&
latency$^\dag$
&
\num{1265056}
&
\num{1416208}
&
\num{1834248}
&
\num{2578812}
&
\num{4093040}
&
\num{6931748}
&
\num{12875024}
&
\num{24729872}
\\
\cline{2-10}
&
key
pair
storage$^\ddag$
&
\num{3321}
&
\num{6642}
&
\num{13284}
&
\num{26568}
&
\num{53136}
&
\num{106272}
&
\num{212544}
&
\num{425088}
\\
\cline{2-10}
&
memory
footprint$^{\ddag\star}$
&\num{38648}
&\num{45520}
&\num{65176}
&\num{78840}
&\num{106392}
&\num{161216}
&\num{270912}
&\num{490336}
\\
\hline
\hline
\end{tabular}

}
{
\small
$^\dag$
(Haswell
cycle).
$^\ddag$
(Byte).
$^\star$
Benchmark
with
\verb"`valgrind --tool=massif --stacks=yes'".
\\
$^\ast$
The
implementation
uses
the
`jump-div-step'
optimization
in~\cite{BY19:extgcd},
consuming
more
stack
space.
}
\end{table*}

\Paragraph{In-browser handshake macrobenchmarks}\autoref{sec:e2e_experiment:benchmark} described how we developed
\code{tls\_timer}, a dedicated handshake benchmarking client, to measure
the
end-to-end
performance
of
OpenSSLNTRU.
The
need
to
fully
control
the
sampling
process
with
\code{tls\_timer},
arose
after
an
initial
attempt
to
measure
the
end-to-end
performance
from
within
the
GNOME
Web
browser.
Specifically,
we
originally
designed
the
experiment
to
let
the
browser
first
connect
to
a
web
server
via
\stunnel{}
to
retrieve
a
static
HTML
page.
This
in
turn
embedded
JavaScript
code
to
open
and
time
a
number
of
connections
in
parallel
to
further
retrieve
other
resources
from
a
web
server.
We
designed
these
resources
to:
\begin{itemize}
\item have short URI to minimize application data in the client
request,
which
length-wise
is
dominated
by
HTTP
headers
outside
of
our
control;
\item have randomized URI matching a ``rewrite rule'' on the web
server,
mapping
to
the
same
file
on
disk.
This
allows
the
server
to
cache
the
resource
and
skip
repeated
file
system
accesses,
while
preventing
browser
optimizations
to
avoid
downloading
the
same
URI
repeatedly
or
concurrently;
\item be short, comment-only, JavaScript files, to minimize
transferred
application
data
from
the
server,
and,
on
the
browser
side,
the
potential
costs
associated
with
parsing
and
rendering
pipelines.
\end{itemize}

Unfortunately,
this
approach
proved
to
be
unfruitful,
as
the
recorded
measures
were
too
coarse
and
noisy.
This
is
mostly
due
to
the
impossibility
of
completely
disabling
caching
on
the
browser
through
the
JavaScript
API
and
developer
options,
delayed
multiplexing
of
several
HTTP
requests
over
a
single
TLS
connection,
ignored
session
keep-alive
settings,
and,
possibly,
the
effect
of
intentionally
degraded
clock
measurements
when
running
JavaScript
code
fetched
from
a
remote
origin.

\section{OpenSSL: software architecture}
\label{sec:ossl_primer}
\autoref{sec:ossl_patches} details the part of our contributions
consisting
of
a
set
of
patches
that
applies
to
the
source
code
of
\opensslLTS{}. We designed our patches to provide full API and ABI
compatibility
with
binary
distributions
of
\opensslLTS{},
while
transparently
enabling
linking
applications
to
perform
post-quantum
key
exchanges
in
\tlsonethree{}
handshakes.
The
description
of
details
of
our
contribution
relies
on
various
technical
concepts
regarding~\openssl{};
this
appendix
reviews
this
background.

Illustrated
in
\autoref{fig:demo/architecture},
as
a
library
to
build
external
applications,
\openssl{}
is
divided
into
two
software
libraries,
namely
\libcrypto{}
and
\libssl{}.
The
former
provides
cryptographic
primitives
and
a
set
of
utilities
to
handle
cryptographic
objects.
The
latter
implements
support
for
\tlsonethree{} and other protocols, deferring all
cryptographic
operations
and
manipulation
of
cryptographic
objects
to
\libcrypto{}.

Due
to
its
legacy,
\libcrypto{}
exposes
a
wide
programming
interface
to
users
of
the
library,
offering
different
levels
of
abstraction.
Currently,
the
recommended
API
for
external
applications
and
libraries
(including
\libssl{})
to
perform
most
cryptographic
operations
is
the
\EVP{} API.
See
\url{https://www.openssl.org/docs/man1.1.1/man7/evp.html}.

The
\EVP{}
API,
especially
for
public
key
cryptography,
offers
a
high
degree
of
\emph{crypto
agility}.
It
defines
abstract
cryptographic
key
objects,
and
functions
that
operates
on
them,
in
terms
of
generic
operations
(e.g.,
\evpencrypt{}/\evpdecrypt{}).
This
lets
the
\libcrypto{}
framework
pick
the
algorithm
matching
the
key
type
and
the
best
implementation
for
the
application
platform.
Using
the
API
appropriately,
a
developer
can
write
code
that
is
oblivious
to
algorithm
selection.
That
is,
leaving
algorithm
adoption
choices
to
system
policies
in
configuration
files,
or
in
the
creation
of
the
serialized
key
objects
fed
to
\libcrypto{}.

In
this
work,
we
patch
\libssl{}
to
support
the
negotiation
of
KEM
groups
over
\tlsonethree{},
mapping
KEM
operations
over
the
existing
\EVP{} API. The API itself does not include abstractions for the \Encap{}
and
\Decap{}
KEM
primitives.

\clearpage
\section{Artifact Appendix}

\subsection{Abstract}

This
demo
was
\begin{itemize}
\item announced 2020.04.16 on the pqc-forum mailing list,
\item updated 2020.04.23 from \openssl[1.1.1f] to \openssl[1.1.1g],
\item updated 2021.06.08 from \openssl[1.1.1g] to \openssl[1.1.1k],
including
additional
support
for
sntrup857,
\item updated 2021.09.30 from \openssl[1.1.1k] to \openssl[1.1.1l],
alongside
an
update
of
the
instructions
to
use
\stunnel{}~\Ver{5.60}
and
\code{glib-networking}~\Ver{2.60.4},
and
\item updated 2021.11.02
to
cover
usage
of
\code{tls\_timer}
and
suggestions
regarding
its
use
for
experiments,
and
\item updated 2021.12.14 from \openssl[1.1.1l] to \openssl[1.1.1m].
\end{itemize}

Our
patches
work
for
versions
of
\openssl{}
from
\Ver{1.1.1f}
to
\Ver{1.1.1m}.

This
is
a
demo
of
OpenSSLNTRU
web
browsing
taking
just
\ourkeygen{}
Haswell
cycles
to
generate
a
new
one-time
\sntrup[761]
public
key
for
each
\tlsonethree{}
session.
This
demo
uses:
(i)
the
Gnome
web
browser
(client)
and
\stunnel{}
(server)
using
(ii)
a
patched
version
of
\openssl[1.1.1m]
using
(iii)
a
new
\openssl{}
\ENGINE{}
using
(iv)
a
fast
new
\sntrup[761]
library.

The
\tlsonethree{}
integration
in
OpenSSLNTRU
uses
the
same
basic
data
flow
as
the
CECPQ2
experiment
carried
out
by
Google
and
Cloudflare.
Compared
to
the
cryptography
in
CECPQ2,
the
cryptography
in
OpenSSLNTRU
has
a
higher
security
level
and
better
performance.
Furthermore,
OpenSSLNTRU's
new
software
layers
decouple
the
fast-moving
post-quantum
software
ecosystem
from
the
TLS
software
ecosystem.
OpenSSLNTRU
also
supports
a
second
NTRU
Prime
parameter
set,
\sntrup[857], optimizing computation costs at an even higher security level.

\subsection{Artifact check-list (meta-information)}

{\small
\begin{itemize}
\item {\bf How much time is needed to prepare workflow (approximately)?: } 60 min
\item {\bf How much time is needed to complete experiments (approximately)?: } 5--60 min
\item {\bf Publicly available?: } Y
\item {\bf Archived (provide DOI)?: } \url{https://opensslntru.cr.yp.to/demo.html}
\end{itemize}

\subsection{Description}

\subsubsection{How to access}

\url{https://opensslntru.cr.yp.to/demo.html}

\subsubsection{Hardware dependencies}

\begin{enumerate}
\item AVX2 support
\end{enumerate}

\subsubsection{Software dependencies}

\begin{enumerate}
\item Linux
\item \opensslLTS{}
\end{enumerate}

\subsection{Installation}

\url{https://opensslntru.cr.yp.to/demo.html}

\subsection{Evaluation and expected results}

We
claim
the
artifact
at
\url{https://opensslntru.cr.yp.to/demo.html}
reproduces
two
of
the
paper
claims.

\subsubsection{Reaching applications transparently}Following
the
provided
instructions,
the
artifact
allows
to
reproduce
\autoref{sec:e2e:reachapps}.
By
the
end
of
the
demo,
you
should
achieve:
\begin{enumerate}
\item Setting up a TLS server with a custom \tlsonethree{} cipher suite supporting \sntrup{}.
\item Setting up a TLS client with a custom \tlsonethree{} cipher suite supporting \sntrup{}.
\item The user sees this upon issuing the last command
\verb+epiphany https://test761.cr.yp.to+ as listed in the demo instructions.
\end{enumerate}

The
server
side
is
optional,
depending
on
if
you
want
to
talk
to
your
own
webserver
or
\url{https://test761.cr.yp.to}.

\subsubsection{Macrobenchmarks: TLS handshakes}Additionally,
the
last
part
of
the
artifact
covers
the
use
of
\code{tls\_timer} to measure the wall-clock execution time of
sequential
TLS
connections
using
different
TLS
groups,
as
described
in
\autoref{sec:e2e_experiment:benchmark}.

Using
\code{tls\_timer}
you
can
evaluate
that

\begin{enumerate}
\item our specialized batch implementation (provided via \engntru{}
by
\libsntrup{761})
is
faster
than
the
reference
code
included
in
the
optional
patch
to
embed
support
for
\sntrup[761]
operations
in
\libcrypto;
\item with the caveats mentioned in
\autoref{sec:e2e_experiment:benchmark}, we achieve new records,
in
terms
of
computational
costs,
when
compared
with
\code{X25519} and \code{NIST P-256}, the fastest pre-quantum
implementations
of
\tlsonethree{}
key-exchange
groups
included
in
\opensslLTS{},
while
providing
higher
pre-quantum
security
levels
and
much
higher
post-quantum
security
levels
against
all
known
attacks.
\end{enumerate}

\subsection{Notes}

To
reproduce
the
results
summarized
in
\autoref{fig:e2e/measures}
in
terms
of
absolute
values,
you
would
have
to
replicate
the
setup
described
in
terms
of
hardware
in
\autoref{footnote:e2e:hw},
and
also
take
care
of
setting
up
both
systems
as
detailed
in
the
paragraph
directly
above
the
footnote
to
avoid
biases
due
to
CPU
contention.
It
should
also
be
noted
that
100
experiments
for
each
group,
each
performing
8192
connections,
will
require
several
hours,
and
that
for
the
entire
duration
of
the
experiment
you
should
ensure
low
network
traffic
and
that
no
other
processes
(automated
updates
and
other
scheduled
processes
in
particular)
are
executed
on
the
machines
running
the
experiments.

A
simpler
alternative,
that
would
prove
consistent
with
the
results
presented
in
\autoref{fig:e2e/measures}
in
terms
of
sorting
the
groups
according
to
the
average
connections
per
second,
but
not
necessarily
in
absolute
values,
would
be
to
install
the
server
side
and
the
client
side
of
the
demo
on
the
same
host,
and
then
use
\code{tls\_timer}
over
the
loopback
interface,
lowering
the
number
of
sequential
connections
(e.g.,
to
1024)
to
reduce
the
execution
time
of
each
experiment.

In
any
case,
it
is
important
to
take
care
of
the
details
listed
in
\autoref{sec:e2e_experiment:benchmark} regarding disabling frequency
scaling,
Turbo
boost,
concurrent
services,
and
scheduled
processes,
isolating
physical
cores
exclusively
to
each
of
the
3
processes
(i.e.,
\code{tls\_timer}, \stunnel{}, and \code{apache2}) involved
in
each
experiment
run,
and
disabling/reducing
logging
to
console
or
files,
in
order
to
minimize
external
causes
of
noise
and
achieve
consistent
results.

\relax

\end{document}